\def\oiiiL49{[\ion{O}{3}] $\lambda4959$}
\def\oiiiL50{[\ion{O}{3}] $\lambda5007$}
\def\oiiiL16{\ion{O}{3}] $\lambda\lambda1661,1666$}

\def\NivL9{\ion{N}{4} $\lambda923$}
\def\NivL17{\ion{N}{4} $\lambda1718$}

\def\ciiiL11{\ion{C}{3}] $\lambda1176$}
\def\ciiiL19{\ion{C}{3}] $\lambda1909$}

\def\HeiiL10{\ion{He}{2} $\lambda1085$} 
\def\HeiiL16{\ion{He}{2} $\lambda1640$} 
\def\HeiiL46{\ion{He}{2} $\lambda4686$}

\documentstyle[aas2pp4, epsf]{article}

\begin{document}


\newcommand{\MSOL}{\mbox{$\:M_{\sun}$}}


\title{Far Ultraviolet Spectra of a Non-Radiative Shock Wave in the Cygnus Loop
\footnote{Based on observations made with the NASA-CNES-CSA Far Ultraviolet
Spectroscopic Explorer. FUSE is operated for NASA by the Johns Hopkins
University under NASA contract NAS5-32985.}
}

\author{John C. Raymond \altaffilmark{2}}

\author{ Parviz Ghavamian \altaffilmark{3}}

\author{ Ravi Sankrit \altaffilmark{4} }

\author{William P. Blair  \altaffilmark{4}}

\centerline{and}

\author{Salvador Curiel  \altaffilmark{5}}

\altaffiltext{2}{Smithsonian Astrophysical Observatory, 60 Garden Street,
Cambridge, MA 02138; jraymond@cfa.harvard.edu}

\altaffiltext{3}{Department of Physics and Astronomy, Rutgers University,
136 Freylinghuysen Rd., Piscataway, NJ  08854; parviz@physics.rutgers.edu}

\altaffiltext{4}{Department of Physics and Astronomy, The Johns Hopkins
University, 34th \& Charles Streets,  Baltimore, MD 21218; ravi@pha.jhu.edu, wpb@pha.jhu.edu}

\altaffiltext{5}{Instituto de Astronom\'{i}a, Universidad Nacional Autonoma de Mexico,
Apartado Postal 70-264, 04510 DF, Mexico; scuriel@astroscu.unam.mx }


\begin{abstract}

Spatial and spectral profiles of O VI emission behind a 
shock wave on the northern edge of the Cygnus Loop were obtained with the FUSE
satellite.  The velocity width of the narrowest O VI profile places
a tight constraint on the electron-ion and ion-ion thermal equilibration
in this 350 $\rm km~s^{-1}$ collisionless shock.  Unlike faster shocks 
in SN1006 and in the heliosphere, this shock brings oxygen ions and 
protons to within a factor of 2.5 of the same temperature.  Comparison
with other shocks suggests that shock speed, rather than Alfv\'{e}n
Mach number, may control the degree of thermal equilibration.

We combine the O VI observations with a low resolution far UV spectrum
from HUT, an H$\alpha$ image and ROSAT PSPC X-ray data
to constrain the pre-shock density and  the structure along
the line-of-sight.  As part of this effort, we model the effects 
of resonance scattering of O VI photons within the shocked gas and   
compute time-dependent ionization models of the X-ray emissivity. 
Resonance scattering affects the O VI intensities at the factor of 2
level, and the  soft spectrum of the X-ray rim can be mostly attributed
to departures from ionization equilibrium.  The pre-shock density is
about twice the canonical value for the Cygnus Loop X-ray emitting shocks.

\end{abstract}


\section{Introduction}

\bigskip
The Cygnus Loop is a nearby, bright supernova remnant (SNR).  Because it is
relatively unreddened, it has been extensively studied in the UV with
IUE, Voyager, HUT, HST and FUSE (e.g. \cite{raymond80}; \cite{blair91}; 
\cite{long92}; \cite{vancura93}; \cite{danforth}; \cite{sankrit00};
\cite{bst} 2002; \cite{sb}).  It is often taken as the prototypical middle-aged SNR,
though it was probably formed by a stellar explosion in a relatively low 
density region of the interstellar medium (ISM) where the progenitor formed a
cavity and a dense shell (e.g. \cite{shull85}; \cite{hester94}; \cite{levenson98}; \cite{miyata} 1999).
In regions where the blast wave encounters the dense shell, radiative 
shocks produce bright optical and UV emission.  Regions where
the blast wave encounters low density gas are seen in X-rays and in the Balmer
lines as non-radiative shocks. 
The distance to the Cygnus Loop is now believed to be 440 pc based on
comparison of proper motions and shock speeds (\cite{shullhip} 1991; \cite{blair99}),
as opposed to the canonical estimate of 770 pc.

\medskip
A non-radiative shock is one that heats plasma to a fairly high temperature,
and that has encountered the plasma so recently that the shocked gas
has had insufficient time to radiatively cool (\cite{mckee}).  The optical and UV emission
arises from a narrow zone just behind the shock where the elements pass
rapidly through successive ionization stages. Collisional excitation causes some
of these atoms and ions to emit photons before they are ionized.  At visible wavelengths, 
non-radiative shocks appear as faint filaments of essentially pure Balmer 
line emission (\cite{cr78}; \cite{ckr80}).

\medskip
Non-radiative shocks present 
unique diagnostics for shock speed and for physical processes in shocks,
because the emission is produced before Coulomb collisions and radiative cooling 
erase the signatures of conditions at the shock.
Thermal equilibration among different particle species is especially
important, as it must be understood for reliable interpretation of electron or 
ion temperatures in terms of the shock speed (\cite{raymond01}).  

\medskip
Several non-radiative shocks along the periphery of the Cygnus Loop
have been studied in an effort to obtain shock parameters and to investigate
the physics of collisionless shocks.  The filaments observed were on the
western edge (\cite{raymond80}; \cite{treffers}) and the northeastern boundary
(\cite{raymond83}; \cite{fesenitoh}; 
\cite{blair99}; \cite{sankrit00}; \cite{sb}).  However, relatively low shock speeds
and incipient radiative cooling made the interpretation of these shocks
ambiguous (\cite{long92}; \cite{hester94}).  Recently, \cite{ghavamian} (2001) studied
a filament on the northern boundary. It was chosen because there was no 
indication of the onset of radiative cooling.  Indeed, the cooling time for the
shock parameters derived below exceeds $10^5$ years, while the flow time through the
region of interest is 500 years.  In addition, its morphology
suggested a higher shock speed, making it easier to resolve the H$\alpha$ profile.

\medskip
Recent studies of SNR shocks have shown electron temperatures far below proton temperatures,
$\rm T_e << T_p$,  in shocks faster than 1000 $\rm km~s^{-1}$ (\cite{raymond95}; \cite{laming96};
\cite{ghavamian} 2001) and ion temperatures approximately mass proportional to the
proton temperatures, $\rm T_{i} \simeq (m_i / m_p) T_p$ (\cite{raymond95}).  This suggests
that the plasma bulk velocity is randomized without efficient
sharing of thermal energy among different particle species.  Slower shocks,
in particular the Cygnus Loop shock observed here, are much closer to
electron-ion thermal equilibration.  \cite{ghavamian} (2001) derived a shock 
speed $\rm V_S$ of 270 to 350 $\rm km~s^{-1}$ and $\rm T_e / T_i$ = 0.70-1.0 
from the H$\alpha$ line profile, the H$\alpha$ to H$\beta$ line ratio 
and the intensity ratio of broad to narrow components.
Temperatures derived from X-ray observations in this region
(\cite{decourchelle} 1997) favor
the upper ends of the allowed ranges, 350 $\rm km~s^{-1}$ and $\rm T_e / T_i$ = 1.0.
One might expect the kinetic temperatures of oxygen and hydrogen to be 
close to equilibrium as well.
 
\medskip
A second aspect of the nature of SNR filaments is 
the 3-dimensional structure of the SNR shock.  This is important for
understanding the nature of the interaction between SNR blastwaves and density
inhomogeneities.  It is also needed to disentangle the pre-shock density, $\rm n_0$, 
from the depth along the line of sight, $L$, for estimates of global SNR parameters.
\cite{hester87} demonstrated
that SNR filaments are tangencies between the line-of-sight and a rippled 
sheet of optically thin emitting gas.  The implied relationship between
surface brightness and Doppler velocity for the diffuse emission between
the tangencies can be exploited to derive a 3D model of the shock 
(e.g., \cite{raymond88}; \cite{danforth}).  An alternative approach is to
compare observables that depend on the line-of-sight depth, L,  and the 
pre-shock density, $n_0$, in different ways.  \cite{szent} separated $\rm n_0$ 
and L in the XA region of the Cygnus Loop by comparing the
intensity of a [Ne V] line, which scales as $\rm n_0 L$, with the X-ray
surface brightness, which scales as $n_0^2 L$.  \cite{patnaude} applied a
similar technique to a recently shocked cloud in the SW Cygnus Loop.

\medskip
A quantitative study of the O VI emission requires consideration of
the effects of resonance scattering, both within the emitting sheet of
gas and in the intervening ISM.
The intrinsic intensity ratio of the resonance doublet of a Li-like ion is 2:1, but 
the 2:1 ratio of opacities implies a larger optical depth for the shorter 
wavelength line.  Thus departures from a 2:1 ratio signal non-negligible
optical depths in the O VI lines (\cite{long92}; \cite{sb}).  \cite{cornett} compared the
optically thin emission in [O III] $\lambda$5007 with UV emission dominated
by the C IV $\lambda$1550 doublet to show that bright filaments in the eastern Cygnus Loop are
optically thick in the C IV lines.  They confirmed that the filaments are
tangencies to the line of sight and that resonance scattering is important
for strong UV lines.

\medskip
This paper presents O VI line profiles obtained with the FUSE satellite at
4 positions behind the filament investigated by \cite{ghavamian} (2001)
along with a low resolution FUV spectrum from the Hopkins Ultraviolet 
Telescope (HUT).
The line widths are used to determine the degree of ion-ion thermal
equilibration.  The spectra yield both the line profile information and the
intensity ratio information to permit both of the approaches described above
to constraining the 3D structure of the shock.  We construct models of the emission
and scattering in the sheet of shocked gas to interpret the spectra.
We find nearly complete equilibration between O and H,
a pre-shock density about twice the value usually quoted for the Cygnus Loop
blastwave, and a depth along the line of sight of about 0.7-1.5 pc.  
We also analyze the ROSAT PSPC X-ray image of the region.  A thin, soft
rim appears behind the filament in ROSAT data.  We construct time-dependent
ionization models to resolve the ambiguity between lower shock speed and
emission from lower ionization states as causes of the soft rim.  We find 
that the soft rim at this location is largely due to strong emission from 
moderate ionization states in the region close to the shock.

\bigskip
\section{Observations and Data Reduction}

\bigskip
\subsection{FUSE Observations}

\bigskip
Spectra were obtained on 15 and 16 June, 2000 with the {\it Far Ultraviolet
Spectroscopic Explorer} satellite (FUSE; \cite{moos}, \cite{sahnow}).  A detailed
discussion of the procedures and uncertainties related to FUSE observations of
SNR filaments is given by  \cite{bst} (2002).  Pointings
were made at the four positions listed in Table 1 and shown in Figure 1.  The
uncertainty in the FUSE blind offset is believed to be about 1\arcsec .  The
image in Figure 1 is a superposition of H$\alpha$ (red) and [O III] $\lambda$5007
(green) images obtained through narrowband filters on the 1.2 m telescope at the 
Fred Lawrence Whipple Observatory on
Mt. Hopkins on Nov. 7, 2000.  Conditions were not photometric, and the seeing
was about 1.6\arcsec  .  The image shows a portion of the H$\alpha$ filament
complex that defines the northern edge of the Cygnus Loop.  Faint H$\alpha$
emission well out in front of the main filaments reveals a portion of
the blastwave in lower density gas.
This is also apparent in the ROSAT image as faint X-ray emission ahead of the
main edge of the remnant.  A portion of the shock making the transition from
non-radiative to radiative appears in [O III] emission at the southeast corner of
the Figure 1.  Very faint, diffuse [O II] and [O III] emission in the northeastern part of
the image may arise from the partially ionized pre-shock gas (\cite{bohigas}).

\medskip
As shown in the inset in Figure 1, the 20\arcsec\/ by 4 \arcsec\/ MDRS aperture
was placed parallel to the H$\alpha$
filament, at a position angle of $317^\circ$ E of N.  The aperture was
placed on the brightest H$\alpha$ filament at the same position observed in the
optical by \cite{ghavamian} (2001).  Subsequent slit positions were parallel
to the first, but displaced 6.5\arcsec , 11\arcsec , and 15.5\arcsec\/ behind the
first.  The last position falls upon a fainter H$\alpha$ filament located about
15\arcsec\/ behind the bright one.  Because of uncertain alignment between the LiF 
and SiC channels, we consider only the LiF data for the MDRS spectra.  

\medskip
Spectra through the 30\arcsec\/ by 30\arcsec\/ LWRS aperture were obtained
simultaneously.  The LWRS aperture was located about 3.5\arcmin\/ to the
northwest along the same filament complex, at the positions shown in Figure 1.
The LWRS sensitivity is higher by the ratio of aperture areas.  Moreover, the 
SiC channels can be used to extend the spectral coverage because the offset
between LiF and SiC channels is relatively small compared to the LWRS aperture
size.  However, the LWRS significantly
degrades the spectral resolution for extended emission.  Especially when
optical depth within the emitting gas or in the intervening ISM affects the
line intensities, it is more difficult to derive reliable estimates of the
intrinsic emission intensities.  Therefore, the LWRS spectra
are less useful for the purposes of this paper, and we discuss them only briefly
below. 

\medskip
The data were processed with CalFUSE 2.0.
We extracted MDRS spectra from the complete data set and also extracted spectra from 
the night portions of the orbits alone.  The night-only extraction greatly reduced 
airglow, but it reduced the signal-to-noise of the O VI profiles so severely that we 
present only the data from the full exposures. For each MDRS pointing we shifted 
the Lif1a and Lif2B spectra to align the O I $\lambda$1039.4 airglow features, then added 
the Lif1a and Lif2b spectra weighted by the effective areas of those channels.  The
residuals around the fit to the dispersion solution for the LiF 2B segment are
around 5 $\rm km~s^{-1}$.  The Lif2B aperture may have drifted relative to the Lif1A
aperture by a fraction of the slit width during the course of the observation,
effectively broadening the aperture by 1 or 2\arcsec \/ (\cite{bst} 2002).  The 
absolute radiometric calibration of FUSE is believed to be good
to 10\% , and the relative calibration between 1032 \AA\/ and 1038 \AA\/ should be
even better.

\medskip
Figures 2a and 2b show the profiles of the O VI 1032.93 \AA\/ and 1037.61 \AA\/
lines at the four positions, with the uppermost profiles corresponding to the
positions closest to the blastwave.  Interstellar absorption in the C II $\lambda$1037.02
and $\rm H_2$ $\lambda \lambda$1037.15, 1038.16 lines strongly affects the 
1037.61 \AA\/ line.  While the interstellar lines lie at velocities of -171, -134 and +158 
$\rm km~s^{-1}$ with respect to the O VI line, we have no means to assess
their effects except to compare the 1032.93 \AA\/ and 1037.61 \AA\/ line profiles.
We assume that these absorption lines are important wherever the ratio
I(1037)/I(1032) falls below its intrinsic value of 0.5.  On that basis,
interstellar absorption significantly attenuates the 1037.61 \AA\/ emission
at velocities below about -60 $\rm km~s^{-1}$ and above about +90 $\rm km~s^{-1}$.

\medskip
Figure 3 shows the profile of O VI $\lambda$1032 from the sum of the LWRS
exposures farther up the filament.  The lower resolution of the LWRS smears
the profile enough to mask the double peak structure if it is present.  The
asymmetry suggests that the diffuse region within the LWRS apertures is redshifted.
The profile is similar to what one would expect from
adding the profiles in Figure 2a and degrading the resolution to that of the LWRS,
though the situation is more complicated because the intensity distribution across the
LWRS aperture also affects the profile.   One novel feature is present, however.
The faint wing at redshifts above 200 $\rm km~s^{-1}$ probably represents an additional
component along the line of sight.  Presumably this is due to the shock front
associated with the faint H$\alpha$ and X-ray emission ahead of the filament
observed by FUSE.  If it's speed is 350 $\rm km~s^{-1}$ like that of the filament
observed by FUSE, the shock must be viewed at about 35$^\circ$ to give the observed
Doppler shift.

Table 2 presents the fluxes of the O VI lines at the four positions.  It should
be kept in mind that the 1037 line is attenuated by $\rm H_2$ and C II interstellar
lines, and that the effect of this absorption depends on the intrinsic line profile.
The dereddened fluxes are based on E(B-V) = 0.08 (\cite{miller}) and the extinction
curve of \cite{cardelli}.  H$\alpha$ fluxes computed by scaling the 
H$\alpha$ brightness in each MDRS aperture in the image in Figure 1 to surface brightness
of a long slit spectrum of \cite{ghavamian} (2001) are given for comparison.  This scaling
should lead to fluxes accurate to about 20\%.

For completeness, Table 2 includes the O VI and C III $\lambda$977 fluxes measured
from the sum of the 4 LWRS spectra. Because of the higher sensitivity of the
LWRS spectra and the summation of 4 spectra, we avoid airglow contamination by using only data 
obtained during orbital night.  The C III $\lambda$977 line is
detected, but variations in the background subtraction limit the accuracy of 
the measured flux to $\pm$40\% .
Upper limits for other lines, such as Ne V] 1136.5 and Ne VI]1005.7 are 2\% the 
intensity of O VI $\lambda$1032.  Upper limits on the S VI $\lambda \lambda$933.4, 944.5
lines are about 6\% of O VI $\lambda$1032 because of the lower sensitivity at the
shorter wavelengths.  The C III line is badly affected by interstellar absorption,
so it is difficult to interpret.  If resonant scattering attenuates the O VI intensity by a factor of
2 (see section 3.3.3), the stringent limits on Ne V] and Ne VI] as compared to 
radiative shock models (\cite{hartigan}) confirm that this is a non-radiative shock.

\bigskip
\subsection{HUT Observations}

\bigskip
The Hopkins Ultraviolet Telescope (HUT; \cite{davidsen}) observed the non-radiative shock
during the Astro-2 mission during March 1995.  The 10\arcsec\/ by 56\arcsec\/ slit was
centered at the position of the first FUSE observation with a position angle
of 315$^\circ$.  The HUT aperture encompasses FUSE positions 1 and 2, and it extends
nearly 3 times as far along the filament.  Essentially all of the time when the Cygnus Loop was observable
fell during orbital day, so the geocoronal background was high.  

The HUT spectra cover the 912-1800 \AA\/ range with 3-4 \AA\/ resolution, and the 
radiometric calibration is accurate to about 10\% (\cite{kruk}).  Only the lines
of He II, C IV and O VI were detected.  The 2-photon continuum of hydrogen, C III $\lambda$977 
and the N V doublet at 1238, 1242 \AA \/ are expected to be strong in
non-radiative shocks, but we could not reliably separate these 
features from the wings of geocoronal Ly$\alpha$ or Ly$\gamma$.

Figure 4 shows the wavelength range that covers the C IV and He II
lines as extracted from exposures totaling 2861 seconds.  
The O VI 1032 and 1037 lines are badly compromised by geocoronal emission 
at this resolution. 
Therefore, we report only the intensities of C IV $\lambda\lambda$1548, 1550
and He II $\lambda$1640.  Comparison of measurements from the first and second
pointings (1311 and 1550 seconds, respectively) indicates that fluctuations in the
background level and the uncertainty in the double Gaussian fit dominate the
uncertainties.  We find intensities  3.20$\pm$0.6 
and 4.38$\pm 0.8 \times 10^{-13}~\rm erg~cm^{-2}~s^{-1}$, respectively.  

The lack of intercombination lines such as N IV] $\lambda$1786 and O IV] $\lambda$1400
confirms the non-radiative nature of the shock. In the models of radiative shocks
these lines are comparable to C IV and He II in brightness (\cite{hartigan}).  The
lack of O IV] provides the stronger constraint, demonstrating that the shocked
gas has not yet cooled to $2 \times 10^5$ K.

\bigskip
\subsection{ROSAT Observations}

\bigskip
ROSAT PSPC observations of the Cygnus Loop have been presented by \cite{levenson99} (1999). 
The filament observed here is about 20\arcmin\/ east the center of their 3506 second
North (E) region exposure.  For
the purposes of this paper, we extracted PSPC spectra of fourteen 25\arcsec\/ by
250\arcsec\/ bands parallel to the filament for comparison with theoretical shock models.
The 25\arcsec\/ width is comparable to the spatial resolution of the PSPC, so the spectra
are not entirely independent.  Single temperature
fits with a \cite{rs77} ionization equilibrium model yield temperatures increasing from
0.14 to 0.2 keV over a distance of about 1\arcmin\/ behind the shock.  However, 
ionization equilibrium is a poor approximation close to the shock.
We therefore fold the spectra predicted by the shock models through the ROSAT effective
area to compute the PSPC count rate as a function of position behind the shock.  The
models and results are discussed in section 3.3.4.   

\bigskip
\section{Analysis}

\bigskip 
\subsection{Resonance Scattering}

\bigskip
Before attempting to interpret the line profiles, we need an estimate of
the optical depths in the O VI lines.  Models based on the code described
by \cite{raymond79} and \cite{cr85} predict optical depths of 0.01 to 0.1
along the flow direction.  The filament observed by FUSE
is viewed nearly edge-on, so the optical depths are much larger.  Because
the emitting sheet of gas is much thicker along the line of sight than
in the flow direction, the main effect of finite optical depth is to 
scatter photons out of the line of sight.  As the optical depth in the
$\lambda$1032 line is twice as large as that in $\lambda$1037, the intensity
of $\lambda$1032 is more strongly affected, and the intensity ratio
I(1032)/I(1037) drops below its intrinsic value of 2:1

\medskip
For a slab geometry and the approximation of single scattering, the
intensity of a line is given by

\begin{equation}
I~~=~~\epsilon~(1~-~e^{-\tau})~
\end{equation}

\noindent
where $\epsilon$ is the ratio of emissivity to opacity.  The factor
$\epsilon$ includes the collision strength, the oscillator strength, the electron
density and factors involving the electron temperature and the O VI
line width.  All these factors are nearly constant in the ionization
zone just behind a non-radiative shock, so $\epsilon$ is nearly
constant.  In that case, the effects of resonance scattering
can be assessed from the ratio

\begin{equation} 
R~\equiv~\frac{ 2~I(1037) - I(1032)} { 2~I(1037)} ~=~ \frac{ 2(1-e^{-\tau_{1037}}) - (1-e^{-\tau_{1032}}) } {2(1-e^{-\tau_{1037}})}
\end{equation} 
 
\noindent
with $\tau_{1032} = 2 \tau_{1037}$.  The ratio varies from 0 for $\tau$=0 to 0.5 for $\tau$=$\infty$.

\medskip
Figure 5 shows the observed ratios R for Positions 1 and 4, and Figure 6 shows
the function given in Equation 2.  The ratios for
Positions 2 and 3 are roughly 0.1, but they are quite noisy.  The solid
parts of the curves are the velocity ranges where the data are reliable.
The dotted portions are regions where C II or $\rm H_2$ interstellar
absorption affects the profile, or where the signal in the
1037 line is too low to give a reliable ratio.

\medskip
Figures 5 and 6 imply optical depths $\tau_{1037}$ of order 1-3 through the emitting
regions of Positions 1 and 4. At Positions 2 and 3 the ratios are smaller, suggesting optical depths
of 0.3 to 0.5. The smaller optical depths are consistent with the morphological indication
that the line of sight is farther from tangency at positions 2 and 3.  In a small portion of the range
in Figure 5a near a velocity of 0 $\rm km~s^{-1}$ R exceeds the 
theoretical maximum of 0.5.  This is presumably due to interstellar absorption, which
scatters $\lambda$1032 photons more effectively than $\lambda$1037
photons, but does not compensate by emitting correspondingly more $\lambda$1032 photons.
However,  Figure 5b does not show a corresponding peak at position 4.
This suggests that a combination of intrinsic and interstellar scattering
is required to account for the peak in Figure 5a.

\bigskip
\subsection{Thermal Equilibration}

\bigskip
The discussion above shows that optical depth can affect the line profiles,
but the rough constancy of R in Figure 5 suggests that
scattering affects the overall intensity more than it distorts the line profile.  Comparison
of the O VI profiles obtained from the the LiF 1A and LiF 2B channels indicates
that the fluctuations in Figure 5 are marginally real.
The most stringent limit on the line width comes from the sharp feature
at $V~=~ +25$ $\rm km ~s^{-1}$ in the profile for Position 4 that arises
from the trailing filament seen in Figure 1.
Based on Figures 5b and 6, the optical depth at the peak of the bright feature 
in the $\lambda$1032 profile at Position 4 is about 1.5, and the flux is attenuated
by about a factor $\rm (1-e^{-\tau})/ \tau$, or about 0.52.  The fluxes at the
apparent half-power points at  -20 $\rm km ~s^{-1}$ and +60 $\rm km ~s^{-1}$,
on the other hand, are attenuated by factors of 0.63 and 1.0, respectively.
Therefore, the line profile has been broadened by attenuation at its peak, and
the apparent Full Width Half Maximum of 120 $\rm km ~s^{-1}$ corresponds
to a 90 $\rm km ~s^{-1}$ intrinsic width.
It is clear from the profiles of Positions 1 to 3 and from the blue wing
of the Position 4 profile that bulk motions also contribute to the
Doppler width. There is no reliable way to assess the bulk velocity  contribution,
however, so we use the observed line width corrected for resonance scattering as 
an upper limit to the oxygen kinetic temperature

\begin{equation}
T_O ~<~ 2.7 \times 10^6 ~\rm K
\end{equation}

\noindent
This can be compared with the proton temperature of $T_p = 1.4 \pm 0.35 \times 10^6~\rm K$
derived from the H$\alpha$ line width by \cite{ghavamian} (2001).  The time
scale for $\rm O^{5+}$ ions to slow by Coulomb collisions corresponds to a distance
of $\sim$ 10-15\arcsec\/ behind the shock for the density range derived below (\cite{spitzer}),
while the O VI emissivity peaks only 1\arcsec\/
behind the shock in the model presented below.  Therefore, the upper limit in
$T_O$ applies to the immediate post-shock region.  Thus the upper limit to the ratio of oxygen to 
hydrogen kinetic temperatures is 2.5 including the uncertainty in the proton temperature.  

For comparison, \cite{raymond00} (2000) and
\cite{mancuso02} (2002) found O VI line widths near 3/4 $\rm V_s$ (FWHM) for 1000 $\rm km~s^{-1}$ shocks
in the solar corona.  The 350 $\rm km~s^{-1}$ Cygnus Loop shock shows a line width below 1/4 $\rm V_s$.
The Alfv\'{e}n Mach number, defined as the shock speed divided by the Alfv\'{e}n
speed, might be a parameter that controls heating in the shock.
In spite of their high speeds, the shocks in the solar
corona had modest Alfv\'{e}n Mach numbers and modest compressions.  The Cygnus
Loop shock has a lower speed, but probably a higher Alfv\'{e}n Mach number.
\cite{berdichevsky} (1997) found He and O kinetic temperatures roughly twice the
mass ratios times the proton temperatures in shocks in the solar wind.  
Few observations of ion temperatures in supernova remnant shocks are available.  Roughly
mass-proportional temperatures, $T_C \sim 12 T_p$ and $T_{He} \sim 4 T_p$, were
found in a 2300 $\rm km ~s^{-1}$ shock in SN1006 (\cite{raymond95}).

\medskip
We find that for this 350 $\rm km~s^{-1}$ shock $T_O \sim T_p \sim T_e$ even though
its Alfv\'{e}n Mach number is fairly high, as opposed to faster
shocks in which $T_O > T_p > T_e$.  On the other hand, fast shocks in the solar wind with modest
Alfv\'{e}n Mach numbers also show $T_O > T_p > T_e$.  This suggests that shock speed rather than Alfv\'{e}n
Mach number may control the thermal equilibration.

\bigskip
\subsection{Structure in 3 Dimensions}

\bigskip
It is not practical to invert the observed H$\alpha$ image and the O VI profiles
to obtain the shape of the shock, because the
effects of resonance scattering are complicated and sensitive to both the
viewing angle and the geometrical structure (e.g., \cite{wood}).  Therefore, we proceed
by constructing a model, computing the line intensities and profiles for the
4 positions, and adjusting the model parameters to optimize the fit.  We
first discuss the overall nature of the structure and its relationship to
the measured profiles, then describe the model, and finally present the
results.

\bigskip
\subsubsection{Model Framework}

\bigskip
The H$\alpha$ filaments clearly represent the projection of a rippled sheet
on to the plane of the sky as proposed by \cite{hester87}.  The surface brightness
of the shock seen face-on is the product of the number of H atoms swept up per
second and the number of photons emitted by an H atom before it is ionized, 
which is given by the ratio of excitation and ionization rates.  

\begin{equation}
I_0~=~ \frac{1}{4 \pi}~n_{H 0}~V_s~\frac{q_{ex}}{q_{ion}} ~~~\rm photons~cm^{-2}~s^{-1}~sr^{-1}
\end{equation}

\noindent
For the nearly edge-on viewing angle of the filaments,
the brightness is increased by a factor sec($\theta$), where $\theta$ is
the angle between the line-of-sight and the shock normal.  As $\theta$ 
approaches 90$^\circ$, the geometrical enhancement is limited to L/$\rm l_{obs}$,
where L is the depth along the line-of-sight and $\rm l_{obs}$ is the
larger of the thickness of the emitting layer and the spatial 
resolution element.  The two nearly parallel
bright filaments covered by the FUSE MDRS slit positions are tangencies
between the line-of-sight and the emitting sheet, and the diffuse
region in between is where the sheet curves away from tangency.

\medskip
An equation similar to Equation 4 describes the O VI intrinsic brightness, but with $n_O$ replacing
$n_{H 0}$ and the excitation and ionization rates for the O VI lines replacing
those of H$\alpha$.  However,
the projection of the O VI brightness is more complicated.  The thickness
of the H$\alpha$ emitting sheet is smaller than the resolution of the 
images, $<~10^{15}~\rm cm$ in the models presented here, and 
the H$\alpha$ line is optically thin.   For the O VI lines, on the other
hand, the emitting sheet is roughly 5\arcsec\/ thick. 
The optical depths discussed above
present an even greater complication.
The line profiles are determined by the emissivity as a function of position,
the line-of-sight component of the bulk velocity of the shocked gas, the
thermal velocity profile of the O VI ions, and the optical depth as a function
of position and velocity.  We adopt an oxygen abundance of 8.73 from
\cite{holweger}.  This is lower than the widely used value of 8.82, but
it is supported by the work of \cite{allende}.  The emitting layer is thin
enough that radiative cooling has no effect, so the abundances of other
elements do not affect the predictions for O VI.

\bigskip
\subsubsection{ O VI Profile Models}

\bigskip
The emissivity and opacity behind a planar shock wave are computed by
the code described in \cite{raymond79} and \cite{cr85}, which has been
updated to allow electron and ion temperatures to evolve separately.
The model parameters are shock speed, pre-shock density, pre-shock
ionization state, pre-shock magnetic field, and the ratio of electron
to ion temperature at the shock.  The magnetic field and pre-shock ionization
state have little effect on the O VI emission and absorption, and they
are fixed at a neutral hydrogen fraction of 0.5 and a field of 3 $\mu$G.
The electron-proton temperature equilibration was investigated by
\cite{ghavamian} (2001), and in the $\rm T_e / T_p~>~0.7$ range they found, the
initial electron-ion temperature ratio also has little effect.  In 
the extreme non-equilibrium case  $\rm T_e / T_p$ = 0.05, the
thickness of the O VI emitting layer is roughly twice as large as it is
for the $\rm T_e / T_p$ = 1 models we adopt.

The model code starts with the Rankine-Hugoniot jump conditions, then computes
the density, velocity, $\rm T_e$ and $\rm T_i$, and the time-dependent ionization state
of the shocked gas.  Because we are interested only in the narrow ionization
zone behind the shock, the density is very close to 4 times the pre-shock 
density, and the velocity in the shock frame is very close to 1/4 the shock
speed throughout.  For nearly complete electron-ion equilibration in the
shock, the temperature is also nearly constant in the emitting region.
Thus the only important parameters are $\rm V_s$ and $\rm n_0$.
The O VI excitation and ionization rates
are fairly weak functions of temperature above $10^6$ K, and the O VI thermal
line width scales as $\rm T_O^{1/2}$.  The thickness of the shock structure
scales as $\rm n_0^{-1}$, the opacity scales as $\rm n_0$ and the local
emissivity as $\rm n_0^{2}$.

The shape of the shock front along the line of sight is parameterized as

\begin{equation}
\rm  Y~~=~~ a~X~~+~~b~sin( \pi  X / c)
\end{equation}

\noindent
where X is the distance along the line of sight and Y is distance in the plane of the sky.
While this form imposes a symmetry on the structure, the data do not warrant a
function having more parameters.  The models used a pixel size of
$6.6 \times 10^{15}$ cm, or 1.0" at 440 pc.  We also assume that
the shock speed and preshock density are constant along the shock front.  Allowing
them to vary would introduce unresolvable ambiguities.  We use the density, 
temperature and O VI concentration from the planar shock models to compute the 
opacity and emissivity in the O VI lines at positions behind the shock.  
The opacity and emissivity are computed at 10 $\rm km~s^{-1}$ resolution
using the O VI line width, shifted by the line-of-sight component of the
bulk speed.  We adopt an oscillator strength $\rm f_{1032}=0.13$ (\cite{morton}),
an excitation rate coefficient $\rm q_{1032}=1.5 \times 10^{-8}~cm^3~s^{-1}$ (\cite{zhang}),
and an ionization rate from \cite{younger} .

The interstellar O VI absorption is an ``external" parameter of the models.
\cite{shelton} found that the galactic O VI column includes a Local Bubble component
of $\rm N_{O VI}~=~ 1.6 \times 10^{13}~cm^{-2}$ and absorbing regions of
$\rm N_{O VI}~=~ 2-7 \times 10^{13}~cm^{-2}$ separated by 400-1300 pc.  The
Cygnus Loop emission will be absorbed by the Local Bubble component and
perhaps by one other absorbing region.  The ISM absorption is taken to
have a temperature of $10^6$ K.  The widths of individual O VI absorption
components may correspond to $3 \times 10^5$, but multiple
components along the line of sight can result in 
typical line widths corresponding to $10^6$ K (\cite{jenkins}).  The Local Bubble 
alone produces a dip of about 15\% at zero velocity.  The central peak in Figure 5a
suggests some interstellar absorption in that it rises above 0.5.  It
is consistent with a total column $\rm N_{O VI}~=~ 7 \times 10^{13}~cm^{-2}$.

The model code computes the surface brightness

\begin{equation}
I(Y,V)~=~\frac{1}{4 \pi} \int \epsilon (X,Y,V) e^{-\tau (X,Y,V)} dX  \rm~~~~photons~cm^{-2}~s^{-1}~sr^{-1}
\end{equation}

\noindent
for the $\lambda$1032 and $\lambda$1037 lines, and we average the profiles over
4\arcsec\/ in Y to compare with the FUSE observations.  Figure 8 shows model predictions
for the $\lambda$1032 profiles at the 4 slit positions in the same format as the 
observed profiles in Figure 2.  As in Table 2, we assume a factor of 2.8 attenuation
due to reddening.  This model assumes $V_s$ = 350 $\rm km~s^{-1}$, a pre-shock density
of 0.5 $\rm cm^{-3}$ and an interstellar column density $\rm N_{O VI}~=~ 0.7 \times 10^{14}~cm^{-2}$.
The shape parameters
of the emitting sheet were a = -0.17, b = 28 and c = 300.  The shape is shown in Figure 7,
but note the order of magnitude difference in the X and Y scales. This model
assumed an oxygen kinetic temperature equal to the proton temperature in the O VI
emitting region.  Figure 9 shows the predictions for the intensity ratio R shown in
Figure 5.  

To examine the effect of changes in the pre-shock density, we computed
the O VI $\lambda$1032 profiles for a model with $\rm n_0=0.3~ cm^{-3}$.
The total number of photons emitted scales directly as $\rm n_0$,
and the profiles in Figure 9 are very close to those in Figure 7a scaled by a factor
of 3/5.  The optical depths in both models are similar, because the density in the
$\rm n_0=0.3~ cm^{-3}$ is smaller, but the thickness of the O VI layer is larger
by the same factor because the cooling length increases.  There are departures of order 10\% from the 3/5 scaling because the
areas extracted for FUSE positions 1 through 4 correspond to slightly different parts 
of the post-shock ionization structure.

To illustrate the effects of optical depth, Figure 10 shows the intrinsic profile 
emitted at position 1 and the profile after the effects of resonant scattering and
interstellar absorption.  The observed profile is attenuated by a further factor
of 2.8 by reddening.  The  dip at line center is largely due to interstellar absorption.
The profile without ISM absorption would be nearly flat between the two peaks.   The
net effect of scattering, which is dominated by scattering within the emitting sheet,
is to reduce the O VI flux by about a factor of 2.

The resonance scattering will also affect the C IV intensity measured
by HUT.  However, the effect may be less severe than for the FUSE observations, 
because the HUT aperture is larger and more of it is filled by diffuse emission that
arises from more face-on portions of the shock.
The relative intensities of the C IV and He II lines are in reasonable
agreement with the predictions of models such as that  described in section 3.3.1 for a 
350 $\rm km~s^{-1}$ shock provided that carbon is depleted by a factor of 2-3
and that resonance scattering (perhaps dominated by the interstellar C IV absorption)
further attenuates the C IV flux by a factor of 2-3.

As a check on the model parameters we compare the H$\alpha$ fluxes at 4 FUSE positions
with the figures in Table 2.  Each neutral hydrogen atom passing through
the shock emits 0.25 H$\alpha$ photons on average, and the shock emits

\begin{equation}
I_0(H \alpha )~=~0.25~n_{H^0}~V_s / 4 \pi  ~~~\rm photons / cm^2~s~sr
\end{equation}

\noindent
and the area within each FUSE position can be computed from Equation 5.  The
computed H$\alpha$ fluxes are 0.019, 0.017, 0.022 and 0.025 $\rm photons ~cm^{-2}~s^{-1}$ for 4\arcsec\/
by 20 \arcsec\/ boxes at
the four positions respectively for $n_{H^0} = 1.0$.  The fluxes at the first and last positions are
extremely sensitive to the exact position, changing by nearly a factor of 2 with shifts of 1\arcsec\/
in the position.  The variation with position roughly matches that in Table 2, but with significant
scatter.  The ratio of the total of the predicted fluxes to the total of the observed H$\alpha$
fluxes is 4.0 $n_{H^0}$.  Thus the predictions match observations for $n_0$=0.5 and a pre-shock
neutral fraction of 0.5.  Calculation of the flux of ionizing photons from such a shock, mainly
$\rm He^0$ $\lambda$584 and $\rm He^+$ $\lambda$304, suggests an upper limit on the pre-shock neutral
fraction of 0.7 (e.g., \cite{smith94} 1994; \cite{ghavam_tycho} 2000).  Comparison of the O VI and H$\alpha$ fluxes in
Table 2 with the model prediction $I_{1032}/I_{H \alpha }$ $\sim$ 12 (and considering the factor
of 2 attenuation of $I_{1032}$ derived above) suggests a neutral fraction near 1.  On the 
other hand, \cite{ghavamian99} used the ratio of the He II $\lambda$4686 line to H$\beta$ in 
this filament to derive an upper limit of 0.2 for the pre-shock neutral fraction.  This 
would require a preshock density $n_0 > 1.25$.  Such a high density would imply an extremely 
small depth along the line of sight to match the ROSAT X-ray brightness (see section 3.3.4), of order 0.1 pc.  This
small depth would increase the curvature of the sheet by a factor of 7, resulting in a very large separation
of the peaks in the O VI profile.  Thus we prefer the parameters $n_0~\sim 0.5$ and a high
neutral fraction.

\bigskip
\subsubsection{Comparison to Observed O VI Profiles}

\bigskip
Figure 8 shows overall qualitative agreement with the intensity and a line shape that
evolves from double peaked near the primary shock toward single peaked where the
emission from the trailing tangency dominates.  As in the observed spectra, the flux
is highest at position 1, declines through positions 2 and 3, then increases again
at position 4.

In detail, there are two discrepancies.  First, the separation of the peaks at 
position 1 is predicted to be only about 85 $\rm km~s^{-1}$, while the observed 
separation is 110 $\rm km~s^{-1}$.  The predicted separations increase to
95 and 105 $\rm km~s^{-1}$ at positions 2 and 3, closer to the observed separations,
but still somewhat smaller.  Second, while the predicted red peak at position 4 is 
closer to zero velocity than at position 3, as observed, the predicted blue peak is 
nearly as strong as the red peak, while the observed blue peak is much fainter.  The
faint blue peak probably results from lower pre-shock density in the region producing
that part of the profile.  The separations of the peaks at positions 1-3 present a
more substantial problem.

The separations of the peaks are partly due to the interstellar
absorption and partly to the line-of-sight velocity components of the shocked gas.
The ISM absorption cannot plausibly be made much broader, but a higher shock speed
or a larger curvature, yielding larger angles between the line of sight and the 
shock front, would increase the separation.  A higher curvature would require a 
compensating density increase to maintain the surface brightness, and $n_0=0.5$.
is already rather high in comparison with X-ray observations discussed below.  A
higher shock speed would be difficult to reconcile with the H$\alpha$ line width
measured by \cite{ghavamian} (2001).

The predicted separations of the peaks could also be increased by relaxing some of the
assumptions inherent to the model.  We have assumed the flow to be perpendicular
to the shock surface, as will occur for a shock driven by gas pressure.  An oblique
magnetic field can introduce a velocity component along the shock surface, but for
the parameters considered here and fields below 10 $\mu$G, this will less than about
1 $\rm km~s^{-1}$ according to the shock jump conditions (\cite{mckee}).
It is also possible to maintain oblique shocks if the
ram pressure of the flow is significant, for instance as occurs in bow shocks.  The steepest
angle in the model sheet is 6$^\circ$, so the parallel component of the post-shock
flow speed should be small, but flows along the shock at $\sim 20 ~\rm km~s^{-1}$ in
the sense expected for oblique shocks could account for the observed peak separations.

Another explanation would involve cosmic ray acceleration.  We have assumed the
the post-shock flow speed of a gasdynamic shock, 3/4 $\rm V_S$.  If a significant
fraction of the energy dissipated in the shock goes into accelerating cosmic rays,
the shock compression is larger than 4, and the post-shock flow speed is larger
than 3/4 $\rm V_S$.  This explanation would require higher shock speeds
to match observed H$\alpha$ profile and X-ray temperature.  If most of the energy
dissipated goes into cosmic rays, a shock several times faster may be needed
(\cite{boulares}). The higher shock speeds would imply a greater distance to the 
Cygnus Loop in order to match measured filament proper motions.  The O VI and
H$\alpha$ surface brightnesses are proportional to the number of particles
swept up per second, so the higher shock speed would imply a smaller density.

\bigskip
\subsubsection{ROSAT Data}

\bigskip
It is important to check the consistency of the derived parameters against the 
ROSAT X-ray images of the Cygnus Loop.  \cite{levenson99} (1999) included the region we observe in their
study of the narrow soft rim in PSPC images.  \cite{miyata} (1999) observed an area just
to the East in a study of elemental abundances and they favored idea that the blastwave has recently
reached the wall of a cavity.  \cite{decourchelle} (1997) studied the region just to the
West, exploring the multiple temperature components needed to explain the brighter,
softer emission regions.  

Figure 11 shows the ROSAT counts extracted in 25\arcsec\/ by 250\arcsec\/ bins parallel to the filament.
The position of the main H$\alpha$ filament is taken as zero, and 440 pc is taken as
the distance to the Cygnus Loop (\cite{blair99}).  Models were computed with the 
\cite{raymond79} code and \cite{allen} abundances.  The curves show model predictions for a 
350 $\rm km~s^{-1}$ shock with preshock densities of 0.2 and 0.3 $\rm cm^{-3}$.  
The X-ray emissivities predicted by the shock model are attenuated by an interstellar column 
$\rm N_H = 1.5 \times 10^{20}~cm^{-2}$
and multiplied by the ROSAT PSPC effective area and exposure time.  A somewhat
higher  $\rm N_H$ is often assumed for the Cygnus Loop, but \cite{decourchelle} (1997)
require $\rm N_H \simeq 1 \times 10^{20}~cm^{-2}$ to match the very soft X-ray
spectra just to the west of the position we observed.  The model predictions have been added to
an assumed background Cygnus Loop emission of 1500 counts estimated from the brightness
ahead of the filament.  The predicted X-ray counts were scaled to line-of-sight depths of 2 and 4 pc, respectively,
for $n_0$ = 0.3 and 0.2 $\rm cm^{-3}$.
The higher pre-shock density $\rm n_0=0.5$ used for the
O VI profile models would produce a curve higher and narrower than the solid curve in
Figure 11, but the spatial resolution of the PSPC would smear it to agree reasonably well
with the measured values if the line of sight depth is scaled to about 0.7 pc.  The
PSPC resolution would smear the 0.2 $\rm cm^{-3}$ predictions to be broader than the observed
points, so we conclude that $n_0 > 0.2 \rm cm^{-3}$.

The post-shock density and temperature are essentially constant,
and the peak in the predicted brightness results from the presence of moderate ionization
stages that emit efficiently in the soft (0.25 keV) part of the ROSAT bandpass before
the gas reaches ionization equilibrium.  This departure from ionization equilibrium
also produces a lower temperature in simple fits to the ROSAT spectrum, accounting
for the gradual rise from kT = 0.14 keV to kT = 0.2 keV over a distance of about
$5 \times 10^{17}$ cm in the fitted temperatures.

The models are planar, so any bending of the surface will
broaden the profile compared with the model profiles. The 0.3-0.5 $\rm cm^{-3}$ models
require a path length comparable to that suggested by the O VI observations,
while a preshock density of 0.2 requires twice that value.  A lower interstellar column,
$\rm N_H = 10^{20}~\rm cm^{-2}$ would bring the $n_0=0.2$ model into agreement
with the O VI model and push the higher density model to a shorter length scale.
We conclude that  $\rm n_0 = 0.3-0.5 cm^{-3}$ and L=0.7-1.5 pc are consistent with
both the FUSE and ROSAT data.  We note, however, that the soft X-ray
rim predicted by the model arises mostly from C, Si and Fe emission lines,
and that these elements may be locked up in grains, as suggested by the C IV flux
measured by HUT.  If they are liberated by
sputtering over a swept-up column of order $10^{18}~\rm cm^{-2}$ (\cite{vancura94}),
the peak will be suppressed.

The soft rim seen in the ROSAT hardness ratio map of the Cygnus Loop was used
by \cite{levenson99} (1999) to support the idea that the Cygnus Loop blastwave 
recently encountered the shell of a cavity created by the progenitor star.  As
\cite{levenson99} (1999) point out, the soft emission in the X-ray bright regions
can result from a slower shock speed in denser material.  They
note that in the non-radiative regions such as the one examined here, the
rim could result either from a higher density when the shock encounters the
shell, or from the non-equilibrium ionization effect illustrated in Figure 11.
We find that non-equilibrium ionization can account for the soft rim in 
this region, but
only if the refractory elements are present in the gas phase without serious
depletion.  We favor a preshock density somewhat higher than the canonical
value of 0.2 $\rm cm^{-3}$.  This may be related to the lag between the outer faint
emission seen in Figure 1 and the filament we observed.  Thus while the
soft rim is solid evidence for recent interaction with dense gas in the 
regions bright in X-rays and optical emission, its interpretation is still
ambiguous in the Balmer-dominated filaments in the northern Cygnus Loop.

\bigskip
\section{Summary}

\bigskip
Far UV observations of the O VI lines in a non-radiative shock in the northern
Cygnus Loop show that the oxygen and proton temperatures  are fairly close
to equilibrium; $\rm T_O / T_p < 2.5$. 
This is consistent with the effective electron-ion equilibration in
this shock inferred by \cite{ghavamian} (2001) and effective ion-ion
equilibration, as opposed to the strong preferential heating
of oxygen measured in faster shocks in the solar wind.

\medskip
The O VI line profiles show clear evidence for resonance scattering, both in the
profile shapes and in the intensity ratio of the $\lambda$1032 and $\lambda$1037
lines.  While there is some uncertainty as to the relative contributions of
interstellar O VI opacity and opacity within the emitting sheet of gas,
optical depths $\tau_{1037} \simeq 0.5 - 2$ are required, with the highest
optical depths in the bright filaments and at small velocities.  The opacity
effects reduce the total line intensity of the 1032 line by roughly a factor of 2.

\medskip
The line shapes and intensities make it possible to estimate the structure of
the emitting sheet along the line of sight, and to infer a pre-shock density.
These estimates can be checked against ROSAT data.  We find a depth along
the line of sight of 0.7 to 1.5 pc and a pre-shock density of 0.3-0.5 $\rm cm^{-3}$.
A shock speed of 350 $\rm km~s^{-1}$ at the upper end of the range allowed by the
H$\alpha$ profile (\cite{ghavamian} 2001) is needed to match the FUSE and ROSAT data.
The above parameters support that conclusion
to some extent, in that the pre-shock density is roughly twice the canonical
value, but part of that difference results from the difference between the
440 pc distance to the Cygnus Loop and the 770 pc distance used in earlier analyses.

\cite{levenson99} (1999) used the soft outer edge of the Cygnus Loop in
ROSAT images to conclude that the blast wave is slowing down due to a very recent
encounter with higher density material.  
The shock model shows that departure from ionization equilibrium can 
contribute to the softness of the ROSAT spectrum in the outer 1-2\arcmin\/ behind
the Balmer line filaments, reducing the need for a density enchancement in this
region.

\bigskip

\acknowledgments

This work was made possible by the dedication of the teams that built and
operate the FUSE satellite.  It was performed under NASA Grant NAG5-9019
to the Smithsonian Astrophysical Observatory and G00-1035X and G01-2052X
to Rutgers University.

\pagebreak

\newpage


\pagebreak

\vspace*{0.00in}
\hspace*{-0.2 in}
\epsfysize=4.000in
\epsfbox{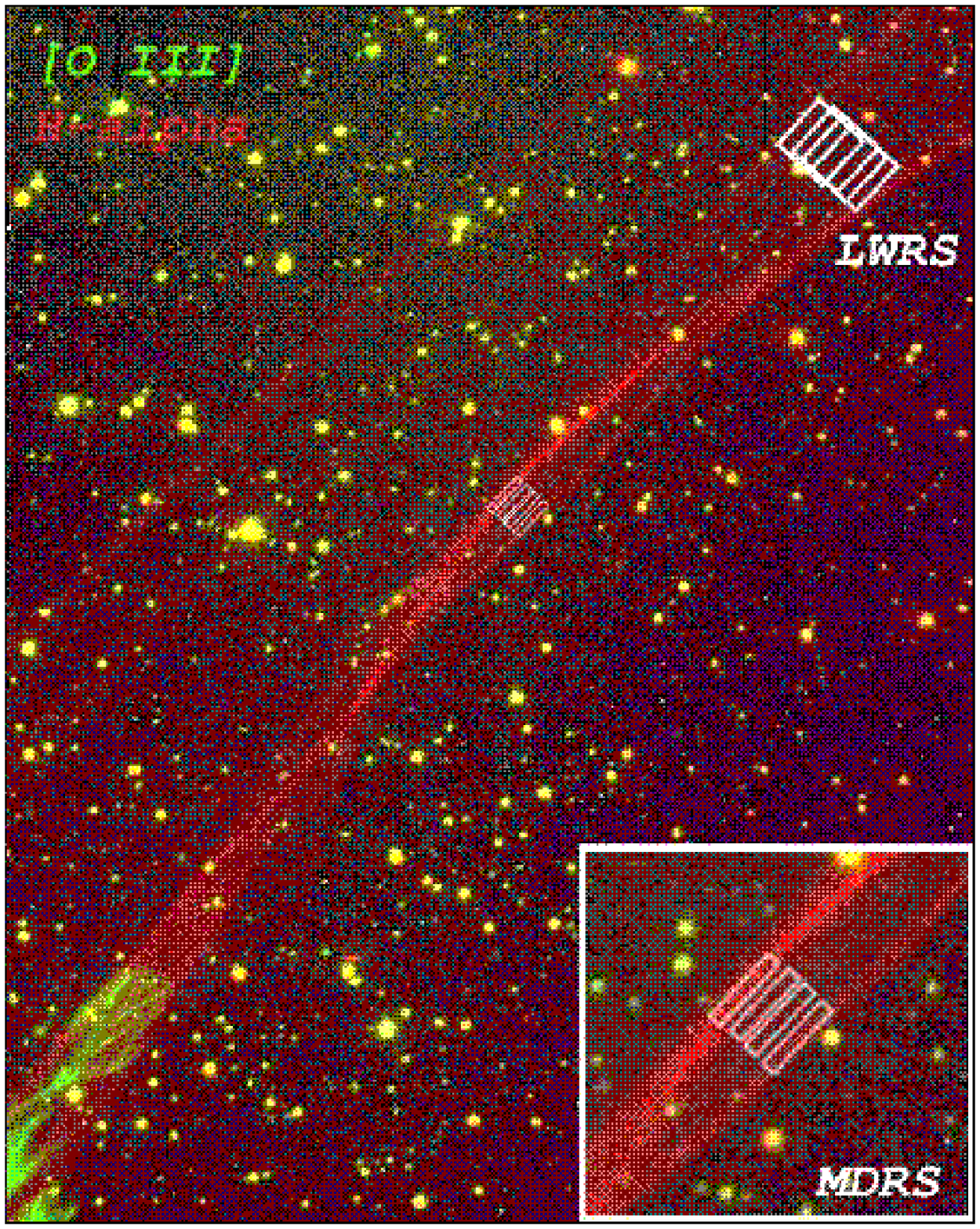}
 
Figure 1:  Images of the northern Cygnus Loop Balmer line filaments in
H$\alpha$ (red) and [O III] $\lambda$5007 (green).  North is at the top
and East at the left.  The image is 7.3\arcmin\/ by 9.2\arcmin .  An expanded
view of the 4 MDRS slit positions is shown in the inset.  The 4 overlapping LWRS
exposures lie on the diffuse emission near the NW corner.   Positions 1 through 4
progress from northeast to southwest.  The 30\arcsec\/ size of the LWRS aperture 
shows the scale.

\pagebreak
\newpage

\hspace*{-0.5 in} 
\epsfysize=3.000in 
\epsfbox{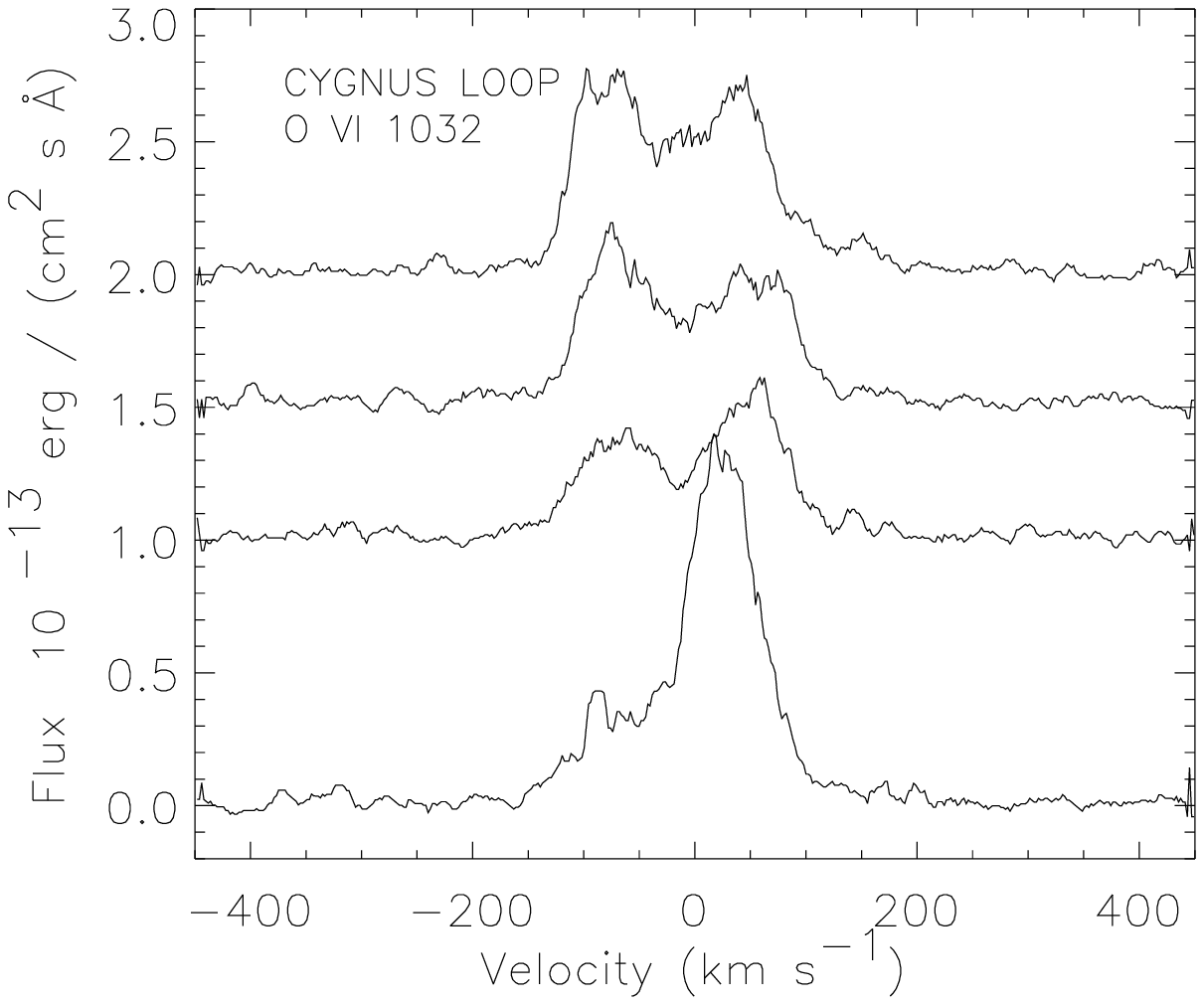}

\hspace*{-0.5 in}  
\epsfysize=3.000in 
\epsfbox{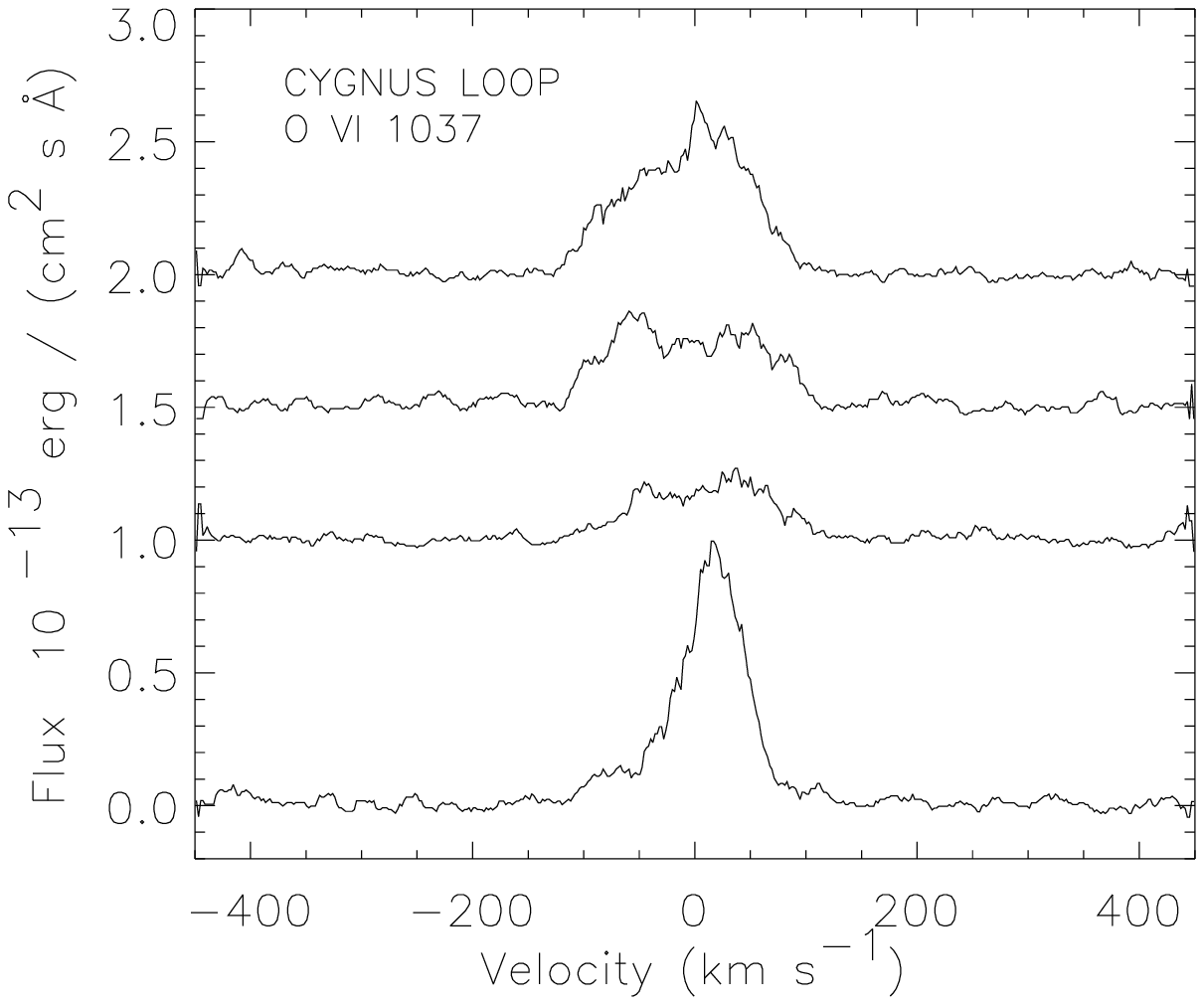} 

Figure 2 -- (a)  O VI $\lambda$1032 line profiles for the four MDRS slit positions, with
the topmost profile corresponding to the leading filament.  Profiles of Positions
1, 2 and 3 are offset by 2.0, 1.5 and 1.0 to avoid overlap.  (b)  Same for O VI $\lambda$1037.
 
\hspace*{-0.5 in}  
\epsfysize=3.000in  
\epsfbox{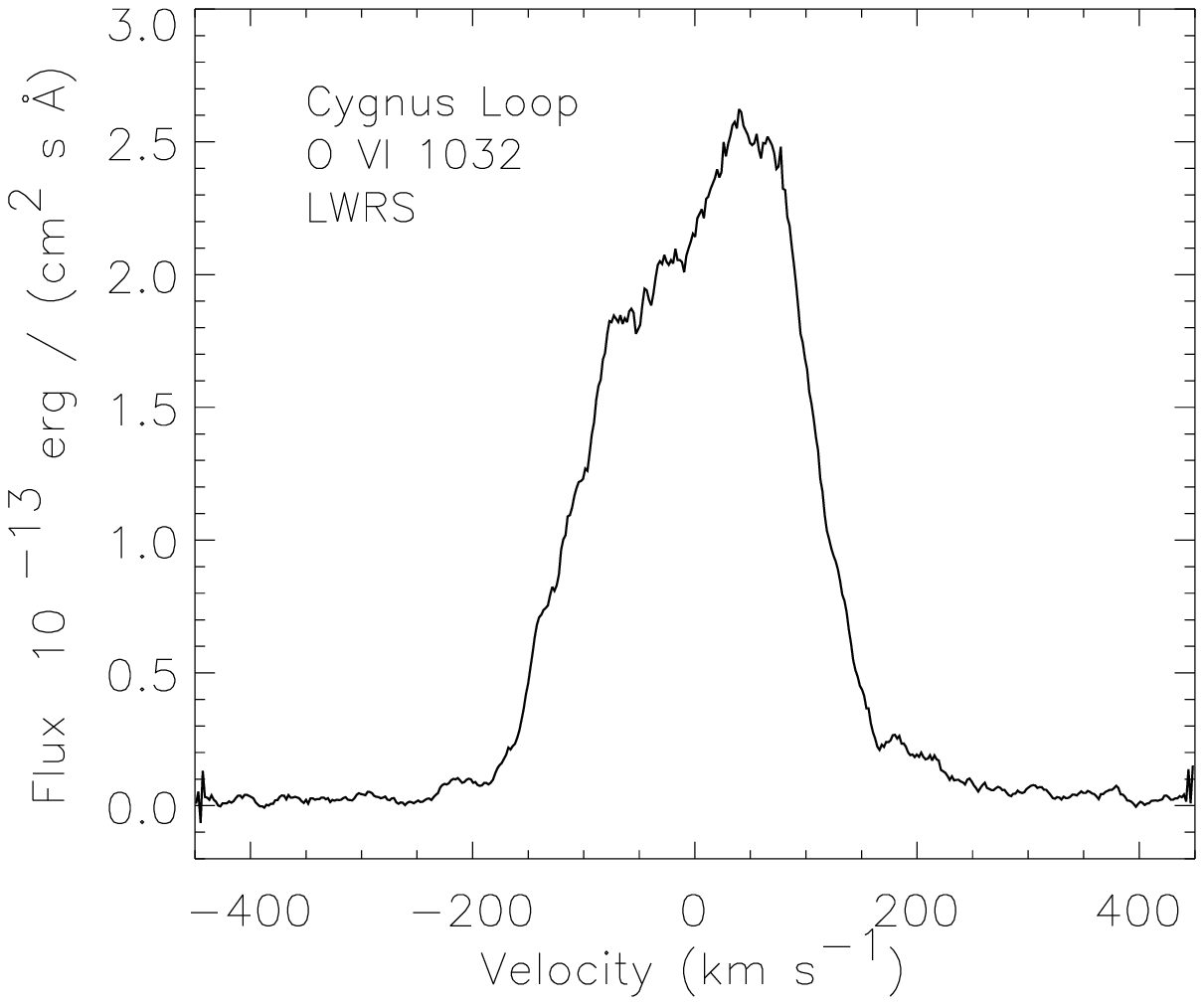}

Figure 3:  O VI  $\lambda$1032 profile for the total nighttime exposure of all four
positions for the LWRS.

\hspace*{-0.5 in}  
\epsfysize=3.000in  
\epsfbox{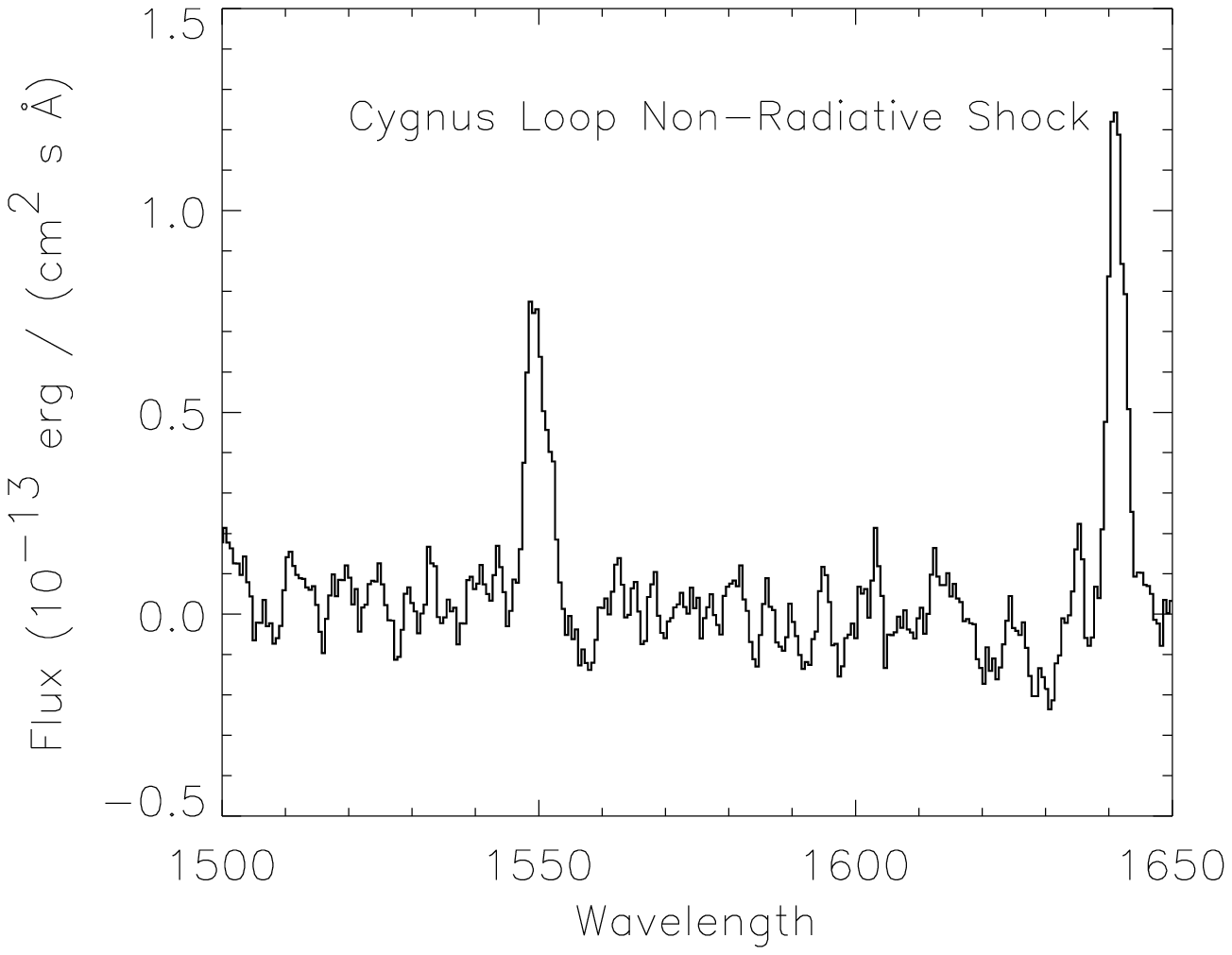}

Figure 4: HUT spectrum showing the wavelength range including 
C IV $\lambda \lambda$1548, 1550 and He II $\lambda$1640.

\hspace*{-0.5 in}  
\epsfysize=3.000in  
\epsfbox{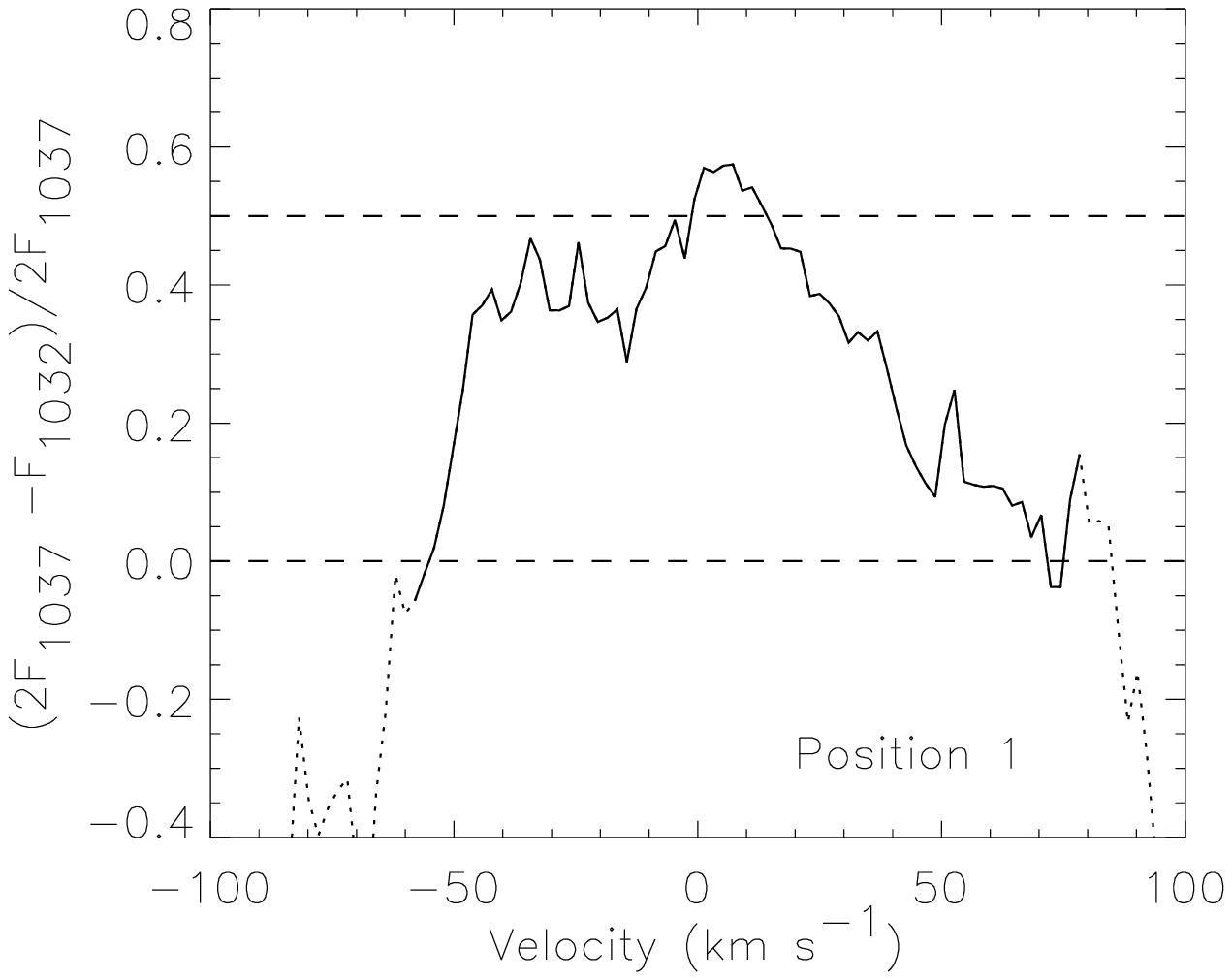}  

\hspace*{-0.5 in}  
\epsfysize=3.000in  
\epsfbox{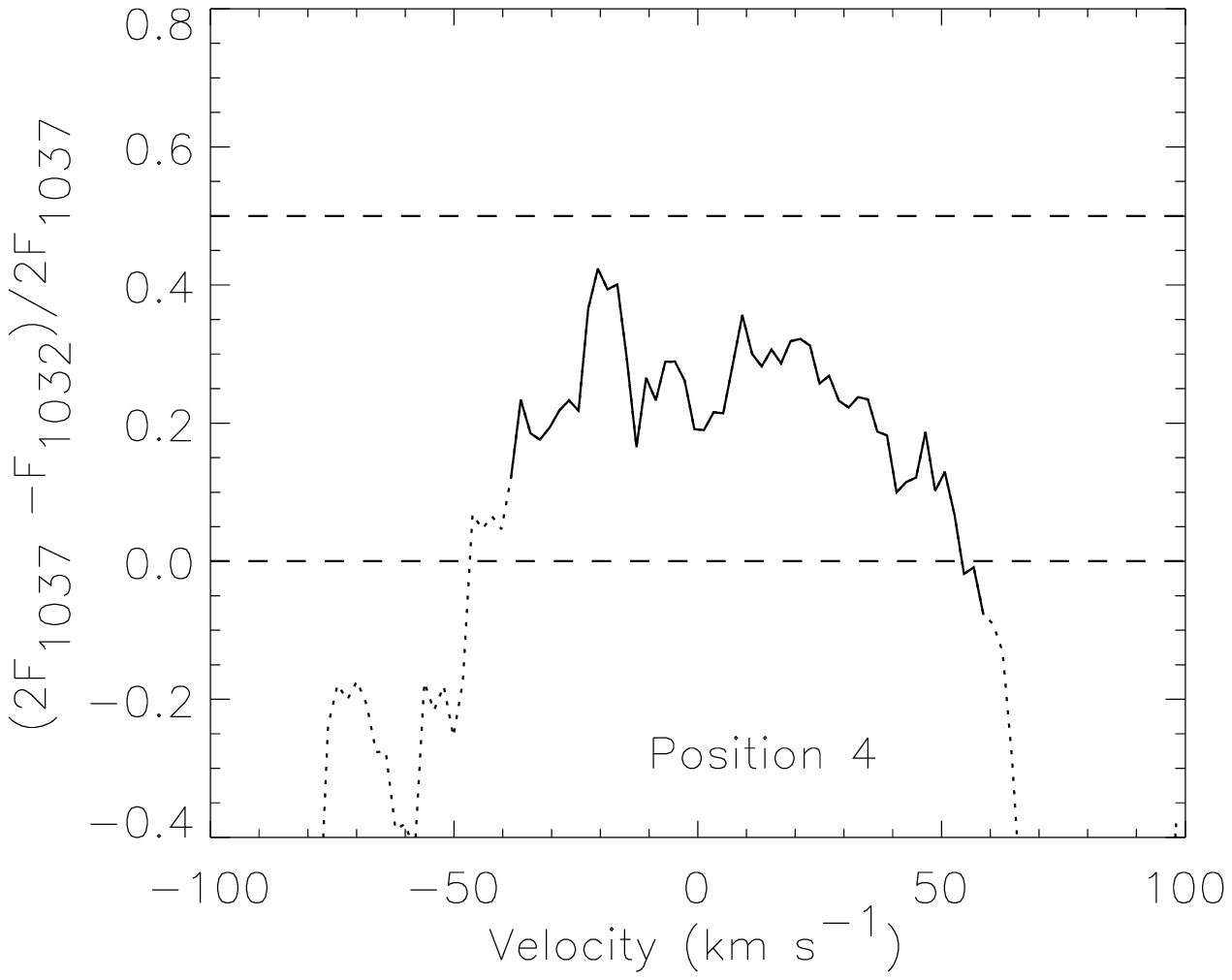}  

Figure 5 --(a)  The intensity ratio of Equation 2 for Position 1.  Dotted line
indicates velocity range where absorption by C II or $\rm H_2$ affects the
ratio. (b)  Same for Position 4.

\hspace*{-0.5 in}  
\epsfysize=3.000in  
\epsfbox{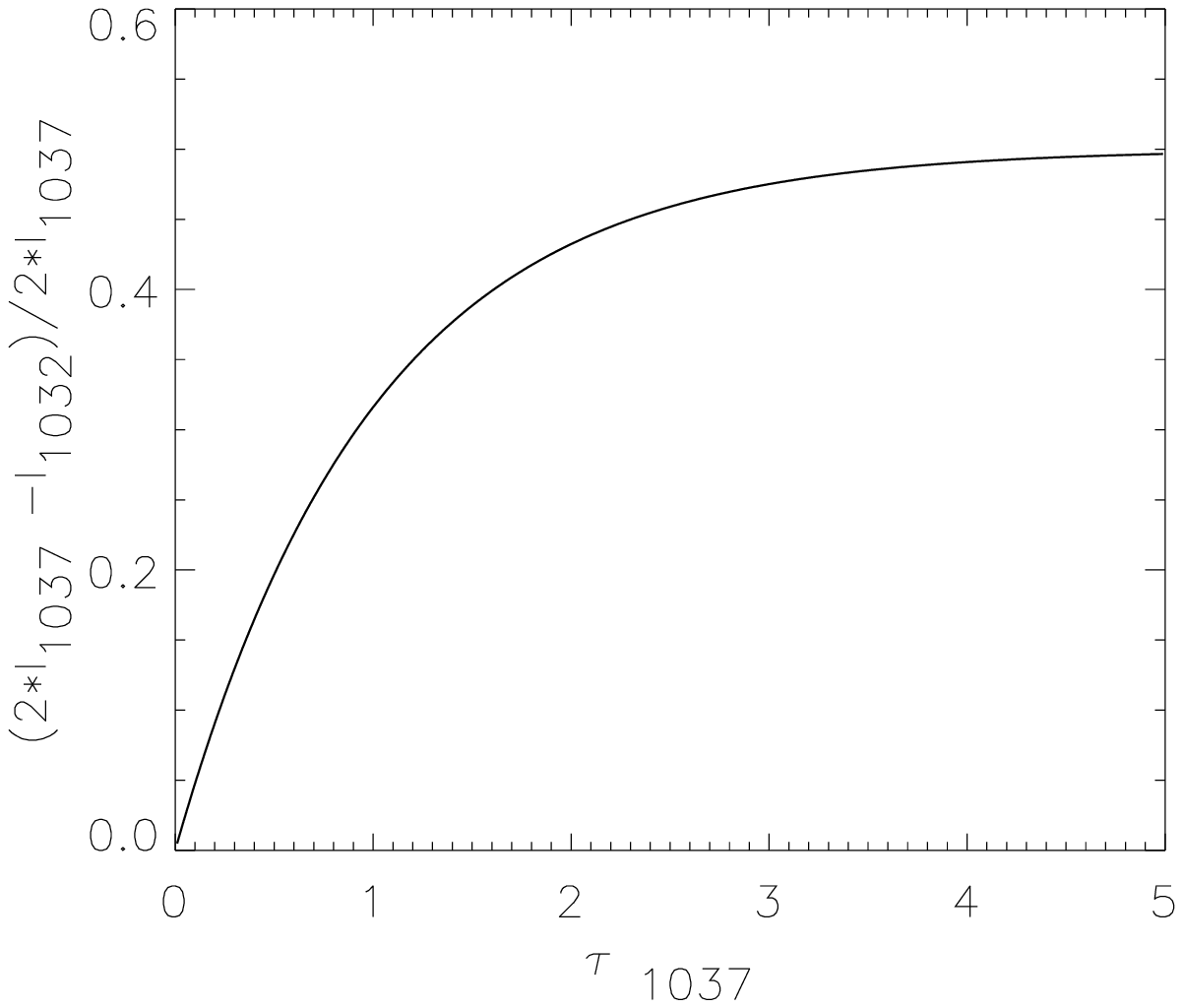}  

Figure 6:  Theoretical value of the intensity ratio R plotted against $\tau_{1037}$.

\hspace*{-0.5 in}  
\epsfysize=3.000in  
\epsfbox{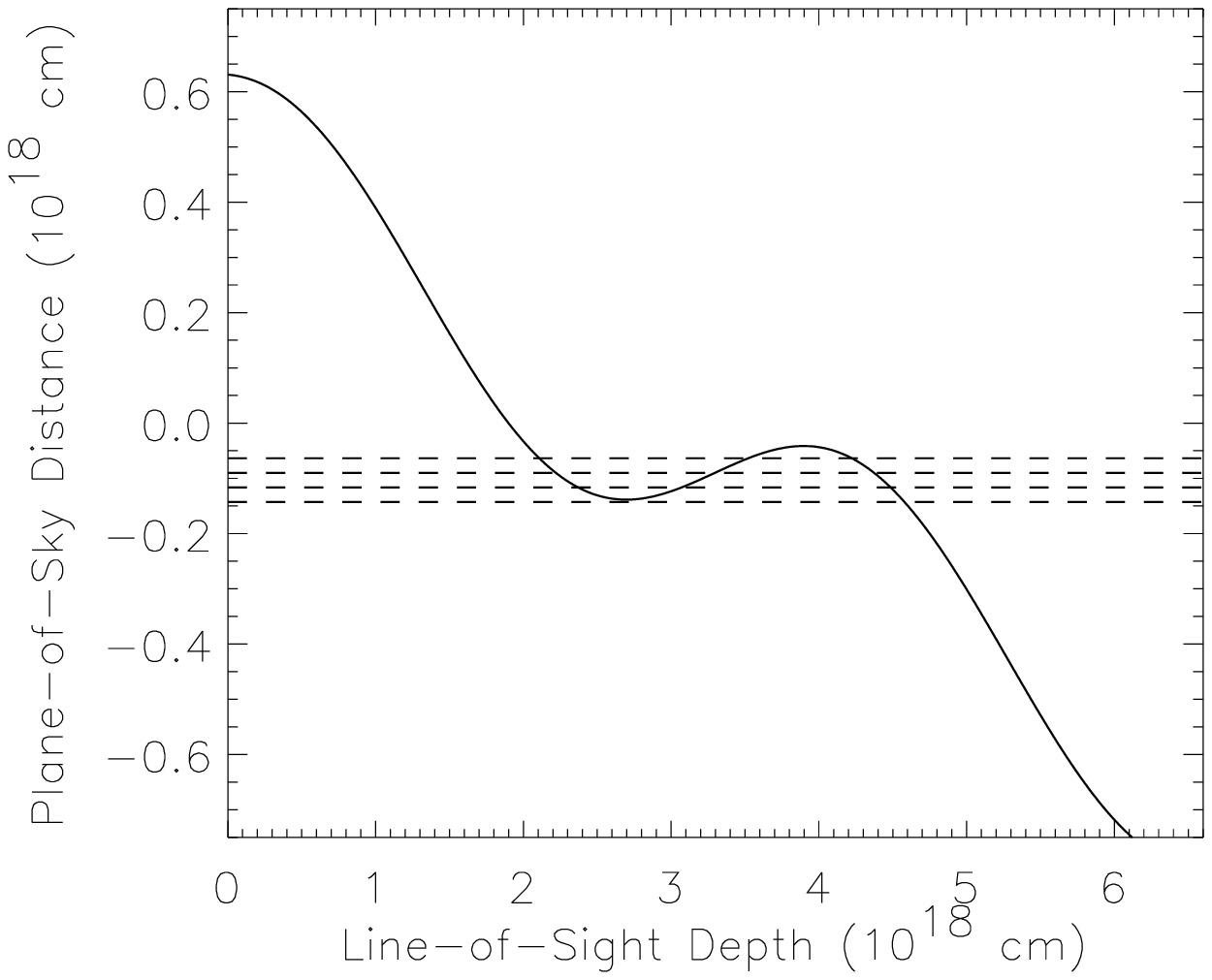}  

Figure 7:  Shape of the emitting sheet along the line of sight.  Note that
the X-scale is compressed by a factor of 5 compared with
the Y scale.  Dashed lines indicate the lines of sight for the 4 FUSE MDRS spectra.

\hspace*{-0.5 in}  
\epsfysize=3.000in  
\epsfbox{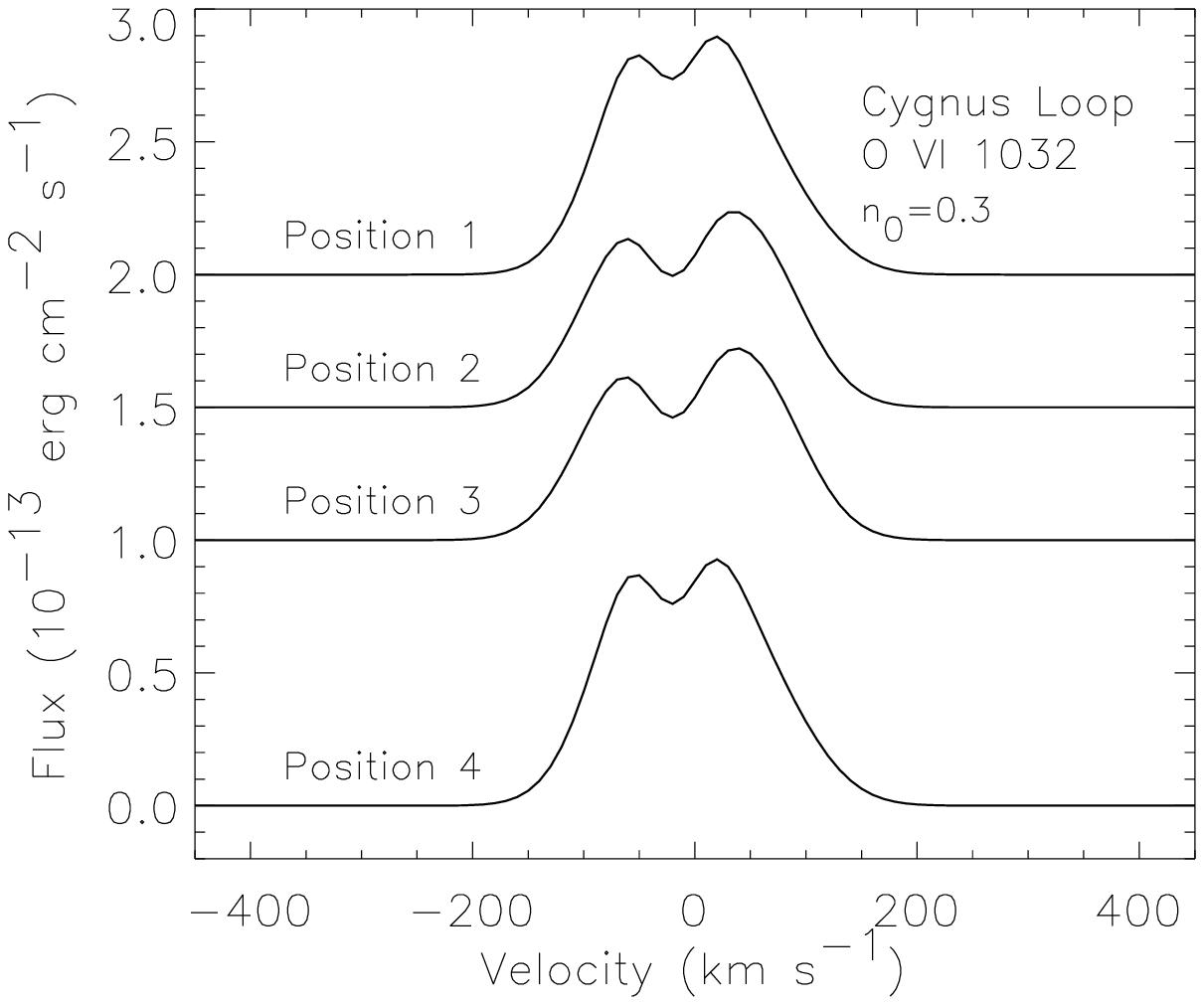}  

\hspace*{-0.5 in}  
\epsfysize=3.000in  
\epsfbox{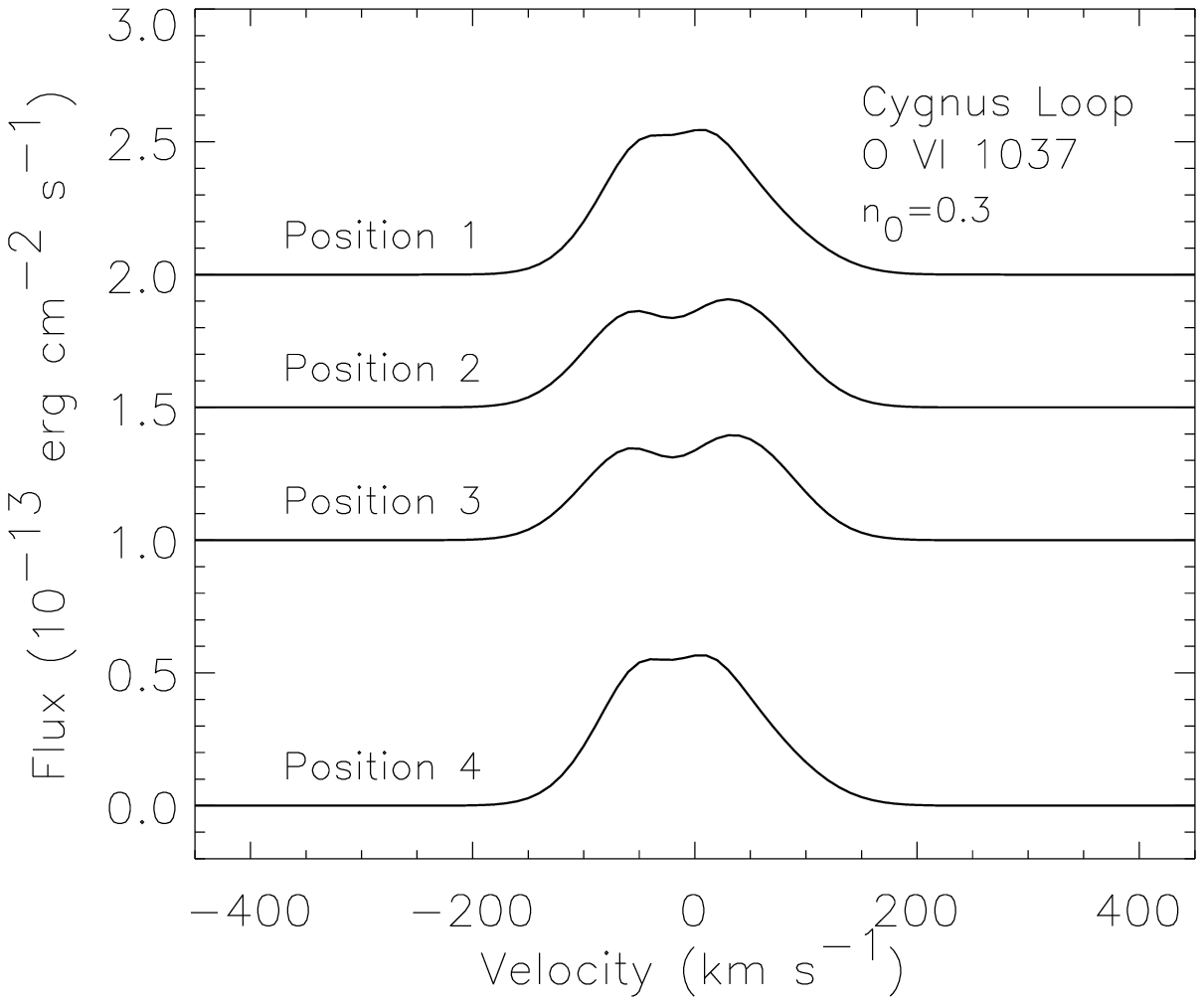}  

Figure 8 --(a)  Predicted O VI $\lambda$1032 line profiles for the four MDRS spectra
for a 350 $\rm km~s^{-1}$ shock with the shape of Figure 7 and a pre-shock
density of 0.5 $\rm cm^{-3}$.  (8b) Same for O VI $\lambda$1037.

\hspace*{-0.5 in}  
\epsfysize=3.000in   
\epsfbox{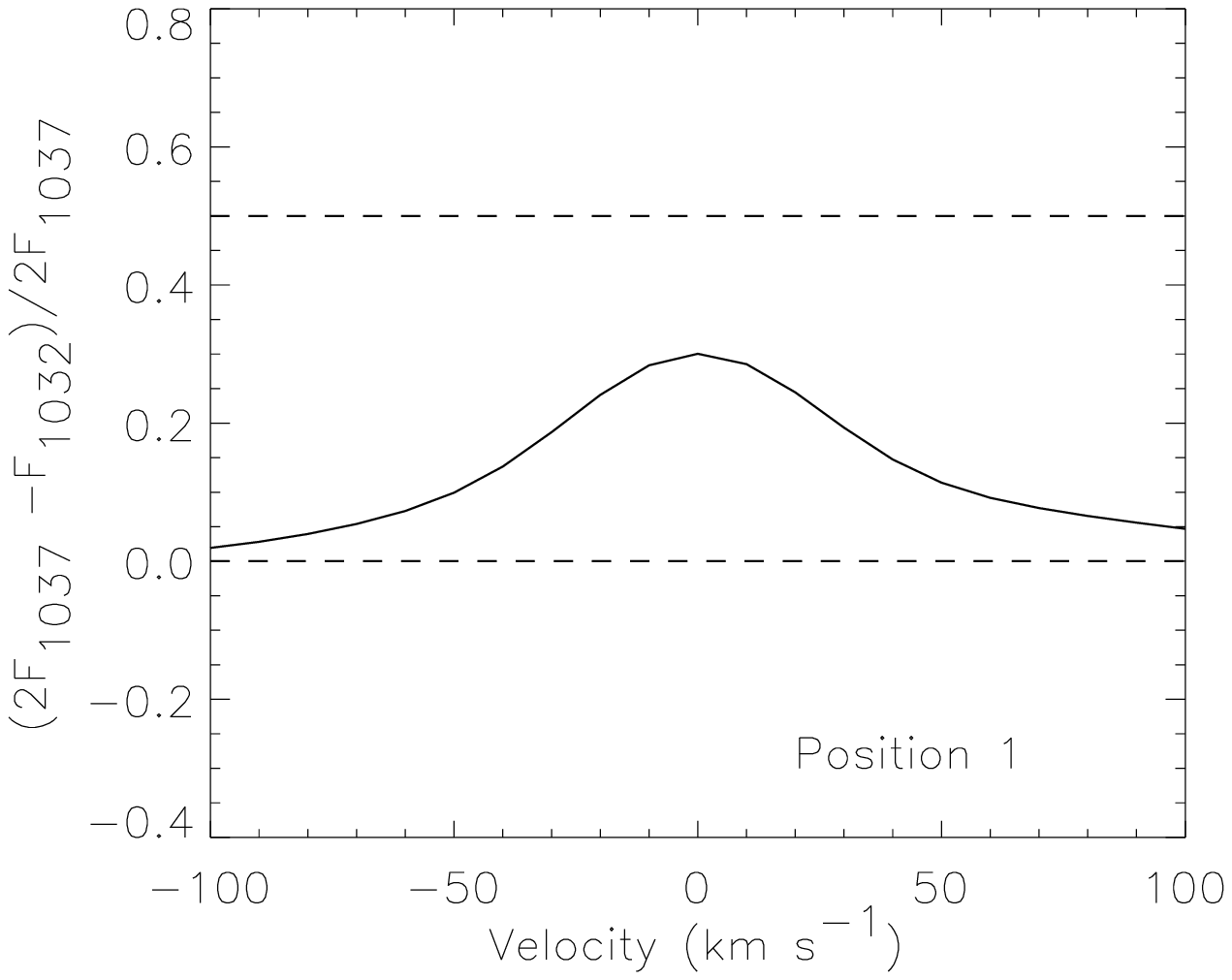}   
 
\hspace*{-0.5 in}   
\epsfysize=3.000in   
\epsfbox{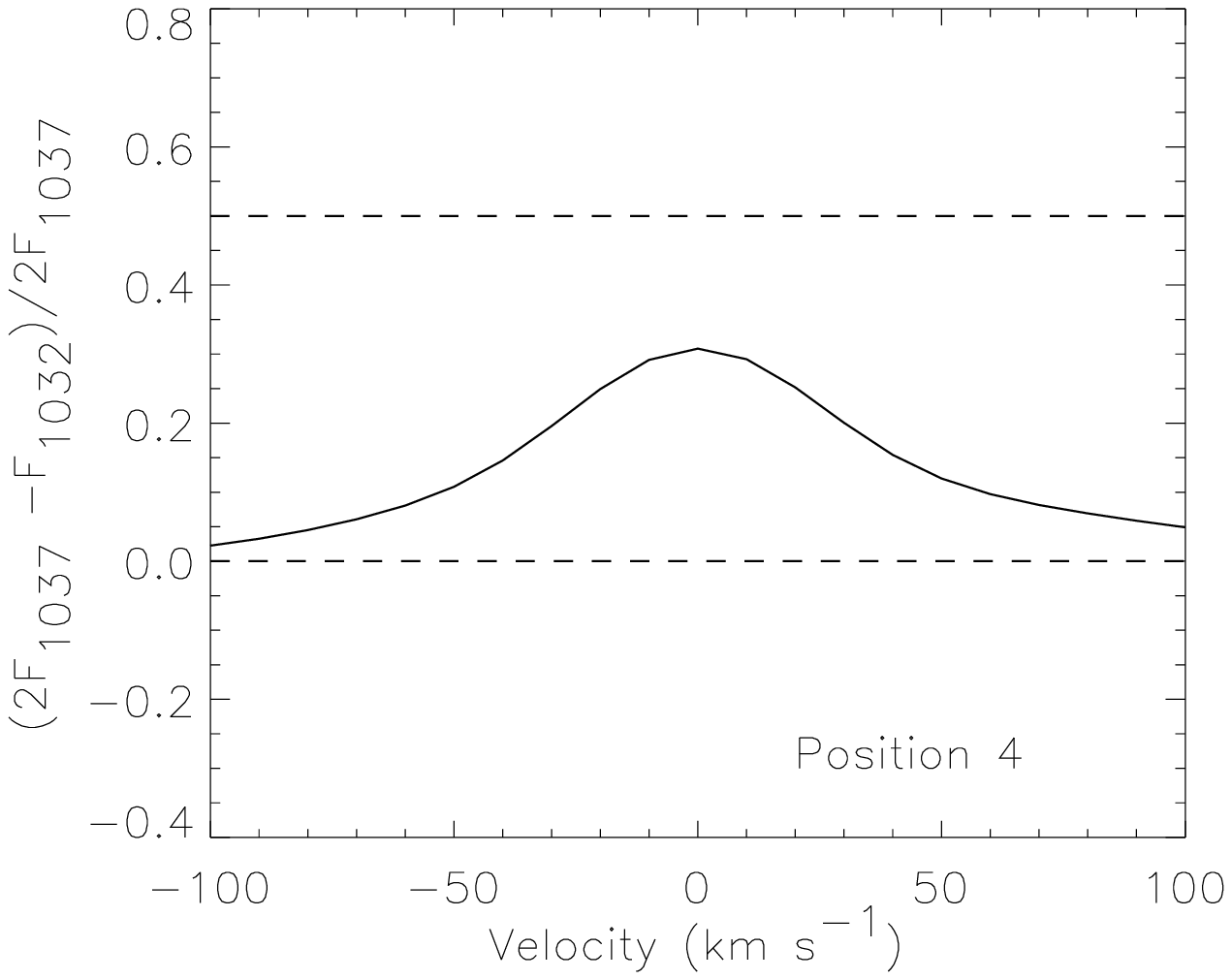}

Figure 9 --(a) Predicted intensity ratios corresponding R to Figure 5a for position 1.
(b) Predicted intensity ratios corresponding to Figure 5b for position 4.

\hspace*{-0.5 in}   
\epsfysize=3.000in    
\epsfbox{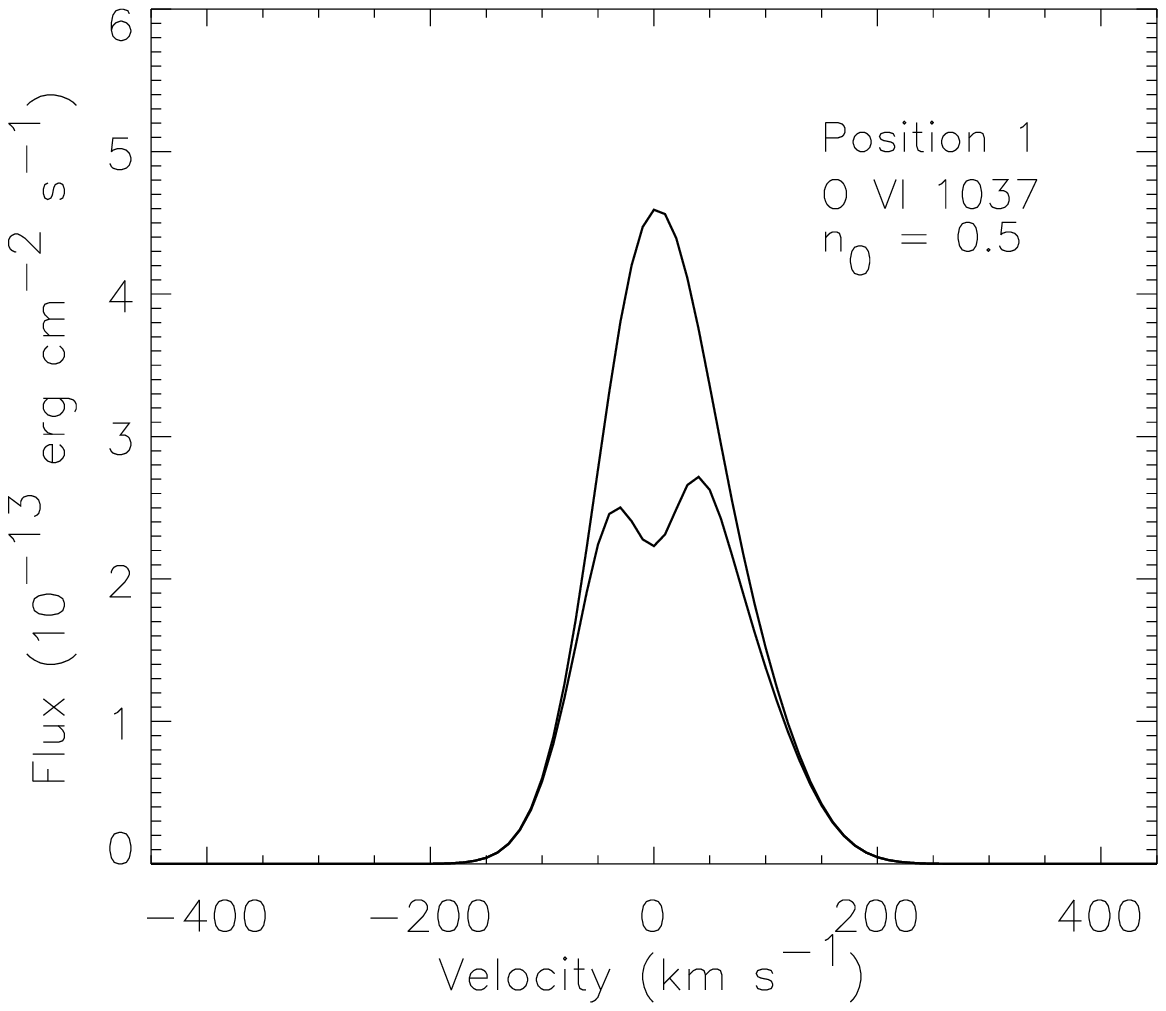}

Figure 10: O VI $\lambda$1032 profile for the optically thin case
and including the effects of interstellar and intrinsic resonance
scattering.  Interstellar reddening will reduce the fluxes by about
a factor of 2.8.

\hspace*{-0.5 in}   
\epsfysize=3.000in    
\epsfbox{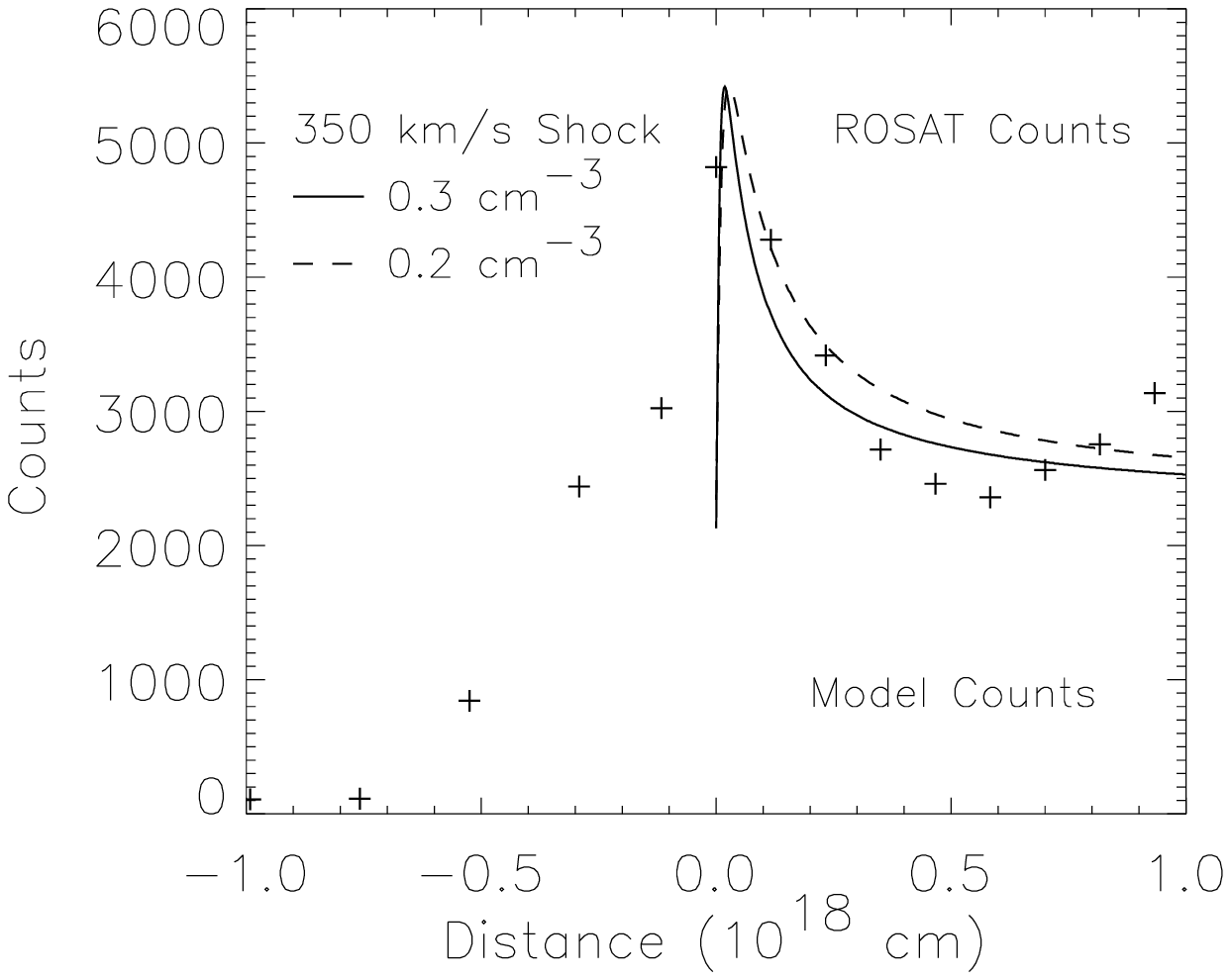}

Figure 11: Predicted ROSAT PSPC count rates as a function of distance behind the H$\alpha$ 
filament,  along with measured values (+).  

\end{document}